\documentclass[9pt,technote,onecolumn]{IEEEtran}

\usepackage{srcltx,epsf,amsmath,amssymb,amsfonts,mathrsfs}
\usepackage{times}
\usepackage{subeqnarray}
\usepackage{epsfig}
\usepackage{url}
\usepackage{cite}
\usepackage{subfigure}

\newtheorem{theorem}{Theorem}
\newtheorem{corollary}{Corollary}
\newtheorem{lemma}{Lemma}
\newtheorem{remark}{Remark}

\def \sinc {{\rm\,sinc\,}}

\begin{document}

\title{On the capacity limit of wireless channels under colored scattering}
\author{Wooseok~Nam,~\IEEEmembership{Member,~IEEE},~Dongwoon~Bai,~\IEEEmembership{Member,~IEEE},~Jungwon~Lee,~\IEEEmembership{Senior~Member,~IEEE},~Inyup~Kang,~\IEEEmembership{Member,~IEEE}\\
\IEEEauthorblockA{Mobile Solutions Lab., Samsung Information Systems America\\
E-mail: wooseok.nam@samsung.com; dongwoon.bai@post.harvard.edu; jungwon@stanfordalumni.org; inyup.kang@samsung.com}
} \maketitle

\begin{abstract}
It has been generally believed that the multiple-input multiple-output (MIMO) channel capacity grows linearly with the size of antenna arrays. In terms of degrees of freedom, linear transmit and receive arrays of length $L$ in a scattering environment of total angular spread $|\Omega|$ asymptotically have $|\Omega| L$ degrees of freedom. In this paper, it is claimed that the linear increase in degrees of freedom may not be attained when scattered electromagnetic fields in the underlying scattering environment are statistically correlated. 
After introducing a model of correlated scattering, which is referred to as the colored scattering model, we derive the number of degrees of freedom. Unlike the uncorrelated case, the number of degrees of freedom in the colored scattering channel is asymptotically limited by $|\Omega| \cdot \min \{ L, 1/\Gamma \}$, where $\Gamma$ is a parameter determining the extent of correlation. In other words, for very large arrays in the colored scattering environment, degrees of freedom can get saturated to an intrinsic limit rather than increasing linearly with the array size.
\end{abstract}


\section{Introduction} \label{SEC:Introduction}

From the time when it was revealed that the multiple-input multiple-output (MIMO) systems promise significant gains, which had never been achievable with any advanced single-antenna systems, in the spectral efficiency and the reliability of communication systems, the MIMO systems have been intensively studied in many different perspectives \cite{Telatar95, Shiu00, Kermoal02, Ozcelik03, Chuah02, Bucci87, Bucci89, Migliore06, Poon05, Raleigh98, Sayeed02, Pollok03, PollokThesis, Browne06, Wallace02, Poon06}. In recent years, along with the advances in implementation technologies, the implementation and deployment of MIMO systems have become active research topics. For example, advanced space-time coding techniques are being adopted by the standards of modern communication systems, such as high-throughput wireless LAN (IEEE 802.11n) and 3GPP Long Term Evolution. In particular, very large MIMO systems \cite{Marzetta10}, also known as massive MIMO, are currently being studied to meet the increasing needs for high-efficiency, green communication systems. Basically, most of these MIMO technologies are based on the premise that the capacity and diversity gains are improved with the increasing number of antennas.

\begin{figure} [t]
\begin{center}
\epsfig{file=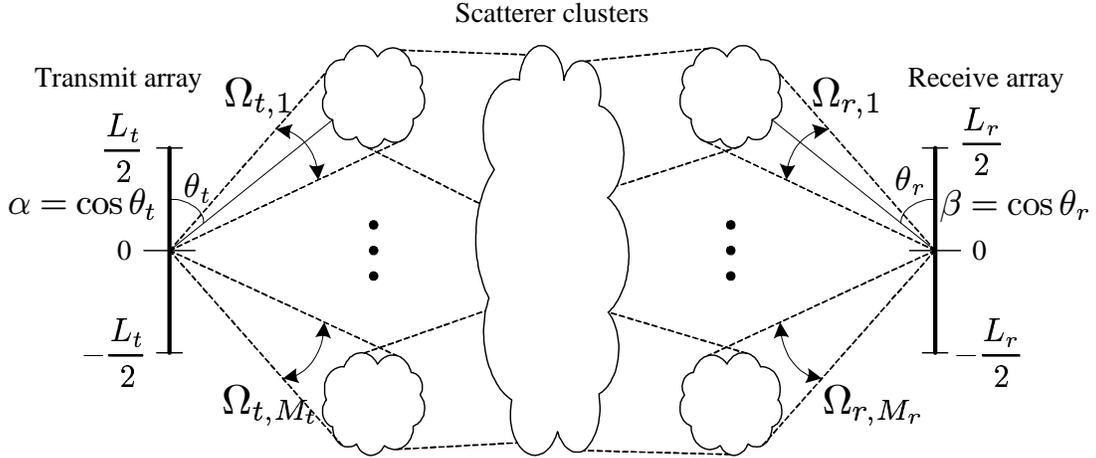, width=0.8\linewidth}
\caption{Clustered scattering model.} \label{FIG:Scattering}
\end{center}
\end{figure}

In a number of early studies on the MIMO fading channel, the rich scattering environment was often assumed and the channel was modeled as a random matrix with independent and identically distributed (i.i.d.) elements, each of which represents the fading channel coefficient between transmit and receive antennas. In the landmark work by Telatar \cite{Telatar95}, it was shown that, for the MIMO channel with i.i.d. fading, the channel capacity grows linearly with the numbers of the transmit and receive antennas. In much subsequent work followed, the i.i.d. MIMO channel was thoroughly studied and lots of insight on the MIMO system was gained. In particular, many space-time coding schemes were designed based on the understanding of i.i.d. MIMO channel model \cite{Tarokh98}. However, in the real applications, the i.i.d. MIMO channel model has very limited importance because the scattering environment is rarely rich enough and the channel coefficients are correlated \cite{Shiu00, Kermoal02}. To address the requirements for more realistic MIMO channel models, the Kronecker model was introduced \cite{Kermoal02, Ozcelik03, Chuah02}. In this model, correlation between channel coefficients is posed by multiplying transmit and receive correlation matrices on the right and left of a matrix of i.i.d. elements. Things become more complicated than the i.i.d. case, but it was shown in \cite{Chuah02} that the capacity of the Kronecker model still grows in proportion to the number of transmit and receive antennas. That is, the correlation introduced in the Kronecker model only impacts the rate of growth of the capacity, not the linearity itself. Though the Kronecker model facilitates the capacity analysis through the powerful random matrix theory, it fails to capture the dependence between the transmit and receive correlation characteristics and has also proven to be inconsistent with measurement results \cite{Ozcelik03}. Recently, efforts were made to incorporate more physical natures of scattering into channel modeling. In the literature related to physical channel modeling \cite{Bucci87, Bucci89, Migliore06, Poon05}, it was stated that, while the transmit and receive signals can be described by the excitation current distribution on antenna arrays ({\em array domain}), the propagation environment can be better explained in the {\em angular domain} in terms of the strength of the electromagnetic fields radiated from or impinging on the arrays. The two domains are convertible to each other through an appropriate transform determined by the geometry of the arrays. In particular, for linear arrays, the mapping between them is given by the Fourier transform, and there exists an analogy between the time-frequency domain and angle-array domain pairs. Given spatial distributions of scatterers, the electromagnetic fields at the transmit and receive sides are related in the angular domain through the {\em Green function}, and the joint response between them, in terms of the angles of arrival and departure, is given by a scattering response function (see Fig. \ref{FIG:Scattering}). In \cite{Raleigh98, Sayeed02}, simplified angular domain channel models were presented based on discrete antennas and discrete scattered paths. Though many insightful observations were made from the discrete path modeling, it is less realistic because the scattered paths are often continuously dispersed and clustered in many practical situations such as urban and indoor environments. Also, for the angular domain channel model, it was shown in \cite{Pollok03, PollokThesis, Browne06} that putting more antennas in a fixed array aperture does not always bring increasing gains due to increased correlation between channel coefficients. Instead, it was observed that the size of the array aperture is more important than the number of antennas and, thus, in \cite{Poon05, Wallace02, Poon06}, the notion of a continuous array, which corresponds to the infinite number of infinitesimally-spaced antennas packed in a given aperture, was adopted. Based on this model, the intrinsic channel characteristics independent of the antenna configuration, such as the capacity and the diversity, were derived as functions of the aperture size and the total angular spread of scattering clusters. In particular, in \cite{Poon06}, three different mechanisms of scattering, i.e., specular reflection, single-bounce diffuse scattering, and multi-bounce diffuse scattering, were introduced and their individual impacts were analyzed.

\begin{figure} [t]
\begin{center}
\subfigure[Low-resolution array]{\epsfig{file=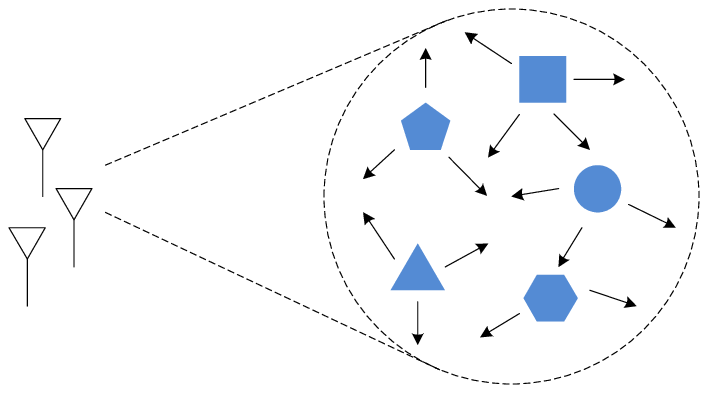, width=0.45\linewidth}\label{FIG:ArrayLowResol}}
\subfigure[High-resolution array]{\epsfig{file=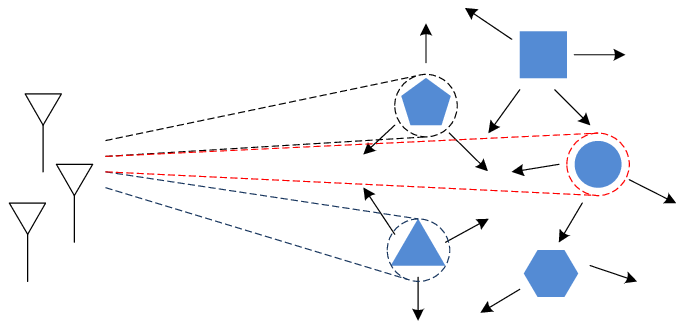, width=0.45\linewidth}\label{FIG:ArrayHighResol}}
\caption{Responses of low- and high-resolution arrays for the same scatterers.} \label{FIG:ArrayResol}
\end{center}
\end{figure}

In this paper, we focus on another intrinsic characteristic of the channel, which has not been addressed in existing literature; the correlation in the scattering response function. The clusters of scatteres are often composed of a large but finite number of scattering objects. Though the scattered fields by a single object can be statistically correlated \cite{Zhang98, ScatteringText}, due to the limited spatial resolution of the arrays, only combined and smoothed effects of the fields that are scattered by multiple scatterers in the resolvable angle are observed as shown in Fig. \ref{FIG:ArrayLowResol}. Therefore, it is reasonable to assume that the scattering response is given by an uncorrelated (or white) Gaussian random process in the angular domain as in \cite{PollokThesis, Poon06}, based on the central limit theorem. However, when we are concerned with a very large array, the spatial resolution can be so high that only a few objects falls into the resolvable angle as in Fig. \ref{FIG:ArrayHighResol}. In such a case, the uncorrelated scattering model seems no longer plausible and some correlation characteristics may need to be appended to the conventional model. For this purpose, we consider a simple correlated scattering model based on the notion of bandlimited white Gaussian random process. For the sake of distinction, we refer to the conventional uncorrelated scattering model in \cite{Poon06} as the white scattering model, and our correlated scattering model as the colored scattering model.

In Section \ref{SEC:ScatteringModel}, we first introduce the colored scattering model and justify it 
compared to the conventional white scattering model. Assuming multi-bounce diffuse scattering, Section \ref{SEC:SpectralDecomp} provides a canonical representation for the colored scattering response by the spectral decomposition technique. Also, Section \ref{SEC:CapBound} shows how the channel is divided into subchannels with linear transmit and receive arrays. The key tools used in Sections \ref{SEC:SpectralDecomp} and \ref{SEC:CapBound} are the Karhunen-Lo\'{e}ve theorem \cite{DavenportText} and prolate spheroidal wave functions \cite{Slepian61}. They have been widely used in the analysis of time- and bandlimited signals, either deterministic or random, and can be directly applied to the analysis of the scattering response due to the analogy. That is, the counterparts of time- and bandlimited signals in the time-frequency domains are the array aperture- or correlation-, and angular spread-limited responses in the angle-array domains. 
Based on the decomposed subchannel model and assuming the fast-fading channel, the ergodic capacity is derived. It is shown that, for a given signal-to-noise ratio (SNR), the pre-log factor of the ergodic capacity, which is also known as the number of degrees of freedom, is bounded by the product of the array length and total angular spread. This is consistent with the previous result for the white scattering model in \cite{Poon06}. However, unlike the previous case, an interesting observation is made for the colored scattering model that the number of degrees of freedom is saturated to a certain value determined by the correlation characteristic of the channel. Similarly in \cite{Raleigh98}, it has already been shown that the degrees of freedom is saturated to the number of discrete and independent propagation paths in the channel. However, there is a fundamental difference between the two saturation phenomena since the scattering response is assumed to be clustered rather than discrete in this paper. To verify the theoretic result derived in Section \ref{SEC:CapBound}, a MIMO system with discrete antennas is considered in Section \ref{SEC:NumericalResult}, and its capacity bounds are numerically computed in the colored scattering environment. The numerical results show that the capacity bounds do not linearly increase with the array size but eventually get saturated to certain values. Finally, Section \ref{SEC:Discussion} discusses extensions of the result of this paper. Though multi-bounce diffuse scattering, linear arrays, and fast-fading are mainly assumed in the analysis of this paper, some extensions to the other situations, such as single-bounce diffuse scattering and slow-fading, are straightforward. Also, some concluding remarks are given in Section \ref{SEC:Discussion}.

The following notations are used throughout this paper. The Dirac delta function is denoted by $\delta(\cdot)$ and, for integers $m_1$ and $m_2$, the Kronecker delta is denoted by $\delta_{m_1, m_2}$. $\sinc x$ denotes the sinc function $(\sin \pi x)/(\pi x)$. Vectors or matrices are denoted by boldface symbols and the superscripts $*$, $T$, and $H$ denote conjugate, transpose, and conjugate transpose, respectively. The determinant and trace of a square matrix ${\bf A}$ are denoted by $\det [{\bf A}]$ and ${\rm Tr}({\bf A})$, respectively, and ${\bf I}$ is the identity matrix of the appropriate size. The sets of real and complex numbers are denoted by $\mathbb{R}$ and $\mathbb{C}$, respectively. For real numbers $a$ and $b$ ($a \leq b$), $[a,b]$ is the closed interval $\left\{ x \in \mathbb{R}| a \leq x \leq b \right\}$. The statistical expectation is denoted by $E\{\cdot\}$ and, for a set $\mathcal{A}$, $|\mathcal{A}|$ denotes the Lebesgue measure. $\log x$, with no indicated base, denotes the logarithm of base 2 and the natural logarithm is denoted by $\ln x$. Finally, $o(\cdot)$ is the standard little-o notation.

\section{Scattering channel and array models} \label{SEC:ScatteringModel}

To focus on the intrinsic characteristics of the channel, we consider the continuous array model as in \cite{Poon05, Wallace02, Poon06}, which yields results not depending on the array configuration, such as the number and positions of the antennas. Thus, the conventional discrete antenna model can be seen as a sampled version of the continuous array model. More specifically, we assume linear arrays of length $L_t$ and $L_r$, which are normalized to the wavelength, at the transmitter and the receiver, respectively. At the transmitter and the receiver, we indicate specific positions on the arrays by scalar coordinate variables $p \in \left[ -L_t/2, L_t/2 \right]$ and $q \in \left[ -L_r/2, L_r/2 \right]$, respectively. Then, for a narrow band transmit signal, which is represented by a current distribution $x(p)$ on the transmit array, the received signal, which is also represented by a current distribution $y(q)$ on the receive array, is given by
\begin{equation}
y(q) = \int_{-L_t/2}^{L_t/2} c (q,p) x(p) dp + z(q) \text, \label{EQ:RxSigArray}
\end{equation}
where $c(q,p)$ is the channel response and $z(q)$ is the zero-mean white Gaussian noise process with an autocorrelation function (ACF) $E \left\{ z(q_1) z^* (q_2) \right\} = \sigma^2 \delta (q_1 - q_2)$ for $q_1, q_2 \in \left[ -L_r/2, L_r/2 \right]$\footnote{For consistency's sake, we try to follow the same signal model and reuse the notation as in \cite{Poon05, Poon06}.}. At the transmitter, the transmit power constraint is imposed as
\begin{equation}
E \left\{ \int_{-L_t/2}^{L_t/2} \left| x (p) \right|^2 dp \right\} \leq P \text. \label{EQ:TxPowConst}
\end{equation}
Physically, the channel response is modeled as a cascade of the transmit array response $a_t (\alpha,p)$, the scattering response $h(\beta,\alpha)$, and the receive array response $a_r (\beta,q)$, i.e.
\begin{equation}
c(q,p) = \int_{-\infty}^{\infty} \int_{-\infty}^{\infty} a_r^* (\beta,q) h(\beta,\alpha) a_t (\alpha,p) d\alpha d\beta \text, \label{EQ:ChResp}
\end{equation}
where $\alpha$ and $\beta$ are directional cosines for the transmit and receive arrays, respectively, as shown in Fig. \ref{FIG:Scattering}. Therefore, $h(\beta,\alpha)$ is physically meaningful only for $\alpha \in [-1,1]$ and $\beta \in [-1,1]$ and vanishes for other values of $\alpha$ and $\beta$. In far-field scattering, where the scatterers are sufficiently far from both the transmitter and the receiver, the array responses of the linear arrays are given by

\begin{align}
a_t (\alpha,p) &= e^{-j 2\pi \alpha p} \text,\\
a_r (\beta,q) &= e^{-j 2\pi \beta q} \text.
\end{align}
Note that, $\alpha$ and $\beta$ are variables in the angular domain, while $p$ and $q$ are variables in the array domain. As mentioned in the introduction, the main benefit of assuming linear arrays is that the mapping between the array and angular domains is done by the Fourier transform, which greatly simplifies the analysis. Hence, an equivalent angular domain representation of the relationship in (\ref{EQ:RxSigArray}) is given by
\begin{equation}
Y (\beta) = \int_{-\infty}^{\infty} \int_{-\infty}^{\infty} L_r \sinc \left( L_r (\beta-\beta') \right) h (\beta',\alpha') \int_{-\infty}^{\infty} L_t \sinc \left( L_t (\alpha'-\alpha) \right) X (\alpha) d \alpha d \alpha' d \beta' + Z(\beta) \text, \label{EQ:RxSigAngle}
\end{equation}
where $X(\alpha)$, $Y(\beta)$, and $Z(\beta)$ are Fourier transform pairs of $x(p)$, $y(q)$, and $z(q)$, respectively. That is,

\begin{align}
X(\alpha) = \int_{-L_t/2}^{L_t/2} x(p) e^{-j 2\pi \alpha p} dp \text, \label{EQ:TxSigAngle}\\
Y(\beta) = \int_{-L_r/2}^{L_r/2} y(q) e^{-j 2\pi \beta q} dq \text,
\end{align}
and
\begin{equation}
Z(\beta) = \int_{-L_r/2}^{L_r/2} z(q) e^{-j 2\pi \beta q} dq \text. \label{EQ:NoiseAngle}
\end{equation}
In (\ref{EQ:RxSigAngle}), we can observe the convolution operations with $L_r \sinc (L_r \beta )$ and $L_t \sinc (L_t \alpha )$, which operate as low-pass filters. In other words, the transmit and received signals are smoothed in the angular domain due to the finite apertures of the arrays. As noted in the literature \cite{Sayeed02}, \cite[Ch. 7]{TseText}, those smoothing operations are closely related to the notion of spatial resolution of the arrays. As the array sizes $L_t$ and $L_r$ grows, the smoothing functions have narrower main lobes, which leads to the higher resolution for paths in the angular domain. Note that the convolution between $L_t \sinc (L_t \alpha )$ and $X(\alpha)$ in (\ref{EQ:RxSigAngle}) is redundant because $X(\alpha)$ in (\ref{EQ:TxSigAngle}) is smoothed by construction due to the finite-length transmit array. Nevertheless, we will not leave it out for later convenience.


Now, let us establish the characteristics of the scattering response $h(\beta,\alpha)$. In the clustered scattering environment, the scattering response is modeled as a continuous 
random process. The support of the scattering response, on which the response may have nonzero values, is determined by the placement of scattering clusters. As shown in Fig. \ref{FIG:Scattering}, we suppose that there are $M_t$ and $M_r$ disjoint clusters seen at the transmit and receive arrays, respectively. At the transmit side, the angular subintervals of $\alpha$ subtended by the scattering clusters are denoted by $\Omega_{t,i}$, $i=1,\ldots,M_t$, and the union of the subintervals is defined as $\Omega_t \triangleq \cup_{i=1}^{M_t} \Omega_{t,i}$. Similarly, at the receive side, the angular subintervals of $\beta$ are denoted by $\Omega_{r,i}$, $i=1,\ldots,M_r$, and the union of the subinterval is defined as $\Omega_r \triangleq \cup_{i=1}^{M_r} \Omega_{r,i}$. In \cite{Poon06}, it was shown that the statistical characteristics of the scattering response depend on the scattering mechanisms, such as specular reflection, single-bounce diffuse scattering, and multi-bounce diffuse scattering. In particular, the specular reflection is deterministic in nature, and the responses for diffuse scattering are assumed to be zero-mean white (uncorrelated) random processes with an ACF\footnote{In \cite{Poon06}, scaling of the ACF depending on the propagation distance and the number of bouncing through the path was considered to reflect path loss. Though the path loss can reduce the received signal power, it does not affect the number of degrees of freedom of the channel. Therefore, in our analysis, we omit the scaling factor.}
\begin{equation}
E \left\{ h(\beta_1, \alpha_1) h^* (\beta_2, \alpha_2) \right\} = \delta (\beta_1 - \beta_2) \cdot \delta (\alpha_1 - \alpha_2 ) \text, \label{EQ:WhiteDiff}
\end{equation}
where the support is
\begin{equation}
(\beta_k, \alpha_k) \in \Omega_r \times \Omega_t,\;k=1,2 \text, \label{EQ:MultiDiffSupport}
\end{equation}
for multi-bounce diffuse scattering and
\begin{equation}
(\beta_k, \alpha_k) \in \bigcup_{i=1}^{M_t} \Omega_{r,i} \times \Omega_{t,i}, \; k=1,2\text, \label{EQ:SingleDiffSupport}
\end{equation}
for single-bounce diffuse scattering ($M_t=M_r$). Elsewhere outside of the support, the scattering response is always zero. Though it has been quite often subsumed in the literature \cite{Sayeed02, PollokThesis, Poon06}, the validity of the above white scattering model for diffuse scattering is somewhat suspicious. In \cite{Zhang98, ScatteringText}, the correlation functions of electromagnetic fields scattered by random objects or rough surfaces were studied. The overall correlation function for a scattering cluster can be characterized by superposing the contributions of the constituent objects. However, when there is a huge number of scattering objects in the cluster, finding the exact correlation function for the cluster is almost impractical. Moreover, since the spatial resolution of the arrays is limited in practice, only smoothed effects of those scattered fields are observable as shown in (\ref{EQ:RxSigAngle}) and Fig. \ref{FIG:ArrayLowResol}. Therefore, it seems rather acceptable to assume that the scattering response is a white random process and the observed channel response is Gaussian distributed with zero-mean by the central limit theorem. 
In terms of the number of degrees of freedom of random processes, white random processes have an infinite number of degrees of freedom \cite{Kikkawa88, Kikkawa04}. How many of those degrees of freedom can actually be used is solely determined by the spatial resolution of the array. 
It follows that the number of usable degrees of freedom increases linearly with the array sizes \cite{Bucci89, Poon05, Poon06}. Depending on the composition of the scattering clusters and the array resolution, this may be true to some extent. However, though it could be very large, the number of scattering objects in each cluster is finite in practice. Moreover, as we increase the array size, we may reach a point at which the spatial resolution of the array is comparable to the angular interval subtended by a single scattering object as shown in Fig. \ref{FIG:ArrayHighResol}. At this point, the correlation of scattered fields can be captured by the array and the white scattering model may no longer be reasonable. 
Instead, as a generalization of the white scattering model, we assume that scattering response is a zero-mean proper\footnote{The definition and properties of proper complex random processes can be found in \cite{Neeser93}. In short, the properness states that the real and imaginary parts of the random process are uncorrelated.} Gaussian random process with
\begin{equation}
E \left\{ h(\beta_1, \alpha_1) h^* (\beta_2, \alpha_2) \right\} = \frac{1}{\Gamma_r} \sinc \left( \frac{\beta_1 - \beta_2}{\Gamma_r} \right) \frac{1}{\Gamma_t} \sinc \left( \frac{\alpha_1 - \alpha_2}{\Gamma_t} \right) \text, \label{EQ:ColoredDiff}
\end{equation}
where the support is given by (\ref{EQ:MultiDiffSupport}) or (\ref{EQ:SingleDiffSupport}) depending on whether it is multi-bounce or single-bounce diffuse scattering. The parameters $\Gamma_t$ and $\Gamma_r$ are referred to as the correlation widths at the transmitter and the receiver, respectively, and determines the extent of correlation in the angular domain. In contrast to the ACF of the white random process (\ref{EQ:WhiteDiff}), the ACF (\ref{EQ:ColoredDiff}) corresponds to that of a bandlimited white random process \cite{Kikkawa88}. Note that the ACF (\ref{EQ:ColoredDiff}) approaches to that of the white scattering (\ref{EQ:WhiteDiff}) as $\Gamma_t$ and $\Gamma_r$ tend to zero. Therefore, the new model includes the white scattering model as a special case. In distinction from the white scattering model, we refer to the scattering model with the ACF (\ref{EQ:ColoredDiff}) as {\em the colored scattering model}. For notational simplicity, we denote the ACF (\ref{EQ:ColoredDiff}) as
\begin{equation}
R_h (\beta_1,\beta_2,\alpha_1,\alpha_2) = R_r(\beta_1,\beta_2)R_t(\alpha_1,\alpha_2) \text, \label{EQ:ACFFact}
\end{equation}
where

\begin{align}
R_t (\alpha_1,\alpha_2) &= \Gamma_r \cdot E \left\{ h(\beta, \alpha_1) h^* (\beta, \alpha_2) \right\} \nonumber\\
&= \frac{1}{\Gamma_t} \sinc \left( \frac{\alpha_1 - \alpha_2}{\Gamma_t} \right) \label{EQ:TxCorr}
\end{align}
for $\alpha_1, \alpha_2 \in \Omega_t$ and $\beta \in \Omega_r$ in the multi-bounce case and $\alpha_1, \alpha_2 \in \Omega_{t,i}$ and $\beta \in \Omega_{r,i}$, $i=1,\ldots,M_t$, in the single-bounce case, and

\begin{align}
R_r (\beta_1,\beta_2) &= \Gamma_t \cdot E \left\{ h(\beta_1, \alpha) h^* (\beta_2, \alpha) \right\} \nonumber\\
&= \frac{1}{\Gamma_r} \sinc \left( \frac{\beta_1 - \beta_2}{\Gamma_r} \right) \label{EQ:RxCorr}
\end{align}
for $\alpha \in \Omega_t$ and $\beta_1, \beta_2 \in \Omega_r$ in the multi-bounce case and $\alpha \in \Omega_{t,i}$ and $\beta_1, \beta_2 \in \Omega_{r,i}$, $i=1,\ldots,M_t$, in the single-bounce case. In the remainder of this paper, (\ref{EQ:TxCorr}) and (\ref{EQ:RxCorr}) are referred to as the transmit and receive ACFs, respectively.

\begin{remark}
When there are more than one scattering cluster, it is plausible to assume that the channel responses for different scattering clusters are uncorrelated. For example, for $\alpha_1 \in \Omega_{t,i}$ and $\alpha_2 \in \Omega_{t,k}$, $i \neq k$, we should expect $R_t (\alpha_1, \alpha_2) = 0$. One shortage of the correlation functions (\ref{EQ:TxCorr}) and (\ref{EQ:RxCorr}) is that the uncorrelatedness between scattering clusters is not accounted for in the multi-bounce case. However, when $\Gamma_t$ and $\Gamma_r$ are very small, which is generally the case, and the separation between the clusters is large enough, the ACF (\ref{EQ:ColoredDiff}) does not lose its generality and simplifies the analysis, which follows in the next section.
\end{remark}





\section{Spectral decomposition of the scattering response} \label{SEC:SpectralDecomp}

Through a spectral decomposition with properly chosen basis functions, random processes can be disassembled into uncorrelated random variables providing the canonical representation of the underlying process. In particular, for the scattering response whose ACF is factorized into the transmit and the receive ACFs as in (\ref{EQ:ACFFact}), the spectral decomposition can be applied separately to each side giving the notion of transmit and receive subspaces. The number of effective subspaces with non-vanishing gains and the way that the transmit and receive subspaces are related give us the insight to the intrinsic number of degrees of freedom and the diversity gain of the channel.

It has been discussed in the literature \cite{Slepian61, DavenportText} that bandlimited white random processes with a bounded time support can be decomposed using the prolate spheroidal wave functions (PSWFs) as the basis. From the analogy between the time-frequency domains in waveform channels and the angle-array domains in linear arrays, the PSWFs can also be used for the spectral decomposition of the clustered and colored scattering response. Thus, in this section, we first introduce some properties of the PSWFs and then study the spectral decomposition of the colored scattering response $h(\beta, \alpha)$. In this and the following sections, we are particularly interested in the multi-bounce diffuse scattering whose support is given by (\ref{EQ:MultiDiffSupport}). The extension to the single-bounce case is discussed in Section \ref{SEC:Discussion}.

\subsection{Required notions and theorems}

Before proceeding further, we first introduce some essential notions and theorems. We denote by $\mathcal{L}^2$ the set of all complex valued functions $f(t)$ defined on the real line $\mathbb{R}$ and square integrable, i.e.
\begin{equation}
\int_{-\infty}^{\infty} \left| f(t) \right|^2 dt < \infty \text.
\end{equation}
Analogously, the set of all complex valued functions $f(t)$ defined on $\mathcal{A}$, which is a compact subset of $\mathbb{R}$, and square integrable on $\mathcal{A}$ is denoted by $\mathcal{L}_{\mathcal{A}}^2$. Now, let $R: \mathcal{A}^2 \rightarrow \mathbb{C}$ be a continuous square integrable Hermitian kernel, i.e.
\begin{equation}
R(t,s) = R^* (s,t), \; (t,s) \in \mathcal{A}^2 \text.
\end{equation}
Associated with the kernel, a bounded linear self-adjoint operator $\mathcal{T}_{\mathcal{A}} (R): \mathcal{L}_{\mathcal{A}}^2 \rightarrow \mathcal{L}_{\mathcal{A}}^2$ is defined as
\begin{equation}
\mathcal{T}_{\mathcal{A}} (R) \left[ f(t) \right] = \int_\mathcal{A} R(t,s) f(s) ds \text,
\end{equation}
for a function $f(t) \in \mathcal{L}_{\mathcal{A}}^2$. Things of particular interest to us are the eigenfunctions $\psi_n (t)$ and eigenvalues $\lambda_n$ of the linear operator that satisfy
\begin{equation}
\int_\mathcal{A} R(t,s) \psi_n(s) ds = \lambda_n \psi_n (t) \text.
\end{equation}
In the remainder of this paper, we assume that the eigenfunctions and eigenvalues are correspondingly ordered such that $\lambda_0 \geq \lambda_1 \geq \lambda_2 \geq \cdots$ without loss of generality. Finally, for $W > 0$, a function $f(t) \in \mathcal{L}^2$ is called $W$-bandlimited if
\begin{equation}
\int_{-\infty}^{\infty} W \sinc \left( W (t-s) \right) f(s) ds = f(t) \text.
\end{equation}
In other words, the Fourier transform of a $W$-bandlimited function vanishes outside $[-\frac{W}{2}, \frac{W}{2}]$. Also, the set of all $W$-bandlimited functions is denoted by ${\mathcal B}_W$.




The following theorem is a minor modification of well-known Mercer's theorem \cite{RieszText} and Karhunen-Lo\'{e}ve theorem \cite{DavenportText}, which we refer here without proof.

\begin{theorem} [Mercer \cite{RieszText} and Karhunen-Lo\'{e}ve \cite{DavenportText}]
For a compact interval $\mathcal{A} = [a,b]$ on the real line, let $x(t) \in \mathcal{L}_{\mathcal{A}}^2$ be a zero-mean proper complex random process. The ACF of $x (t)$ is defined as $R ( t, s ) \triangleq E \left\{ x(t) x^*(s) \right\}$, $(t, s) \in {\mathcal A}^2$. We denote the sets of eigenfunctions and eigenvalues of $\mathcal{T}_{\mathcal{A}} (R)$ by $\left\{ \psi_n (t) \right\}$ and $\left\{ \lambda_n \right\}$, respectively. If $R(t, s)$ is strictly positive definite, $\lambda_n >0$, and we can find $\psi_n (t)$ being orthogonal and complete in $\mathcal{L}_{\mathcal{A}}^2$ and
\begin{equation}
\int_{\mathcal A} \psi_{n_1} (t) \psi^*_{n_2} (t) dt = \lambda_{n_1} \delta_{n_1, n_2} \text.
\end{equation}
Then the random process $x(t)$ can be expanded in terms of $\psi_n (t)$ as
\begin{equation}
x (t) = \sum_{n=0}^{\infty} x_n \psi_n (t),\;t \in {\mathcal A} \text, \label{EQ:KLOrthoExp}
\end{equation}
where the convergence is in the mean-square sense\footnote{The convergence in the mean-square sense is differently stated for deterministic and random signals: for $x(t) \in \mathcal{L}_{\mathcal{A}}^2$,
\begin{equation*}
\lim_{N \rightarrow \infty} \int_{-\infty}^{\infty} \left| x(t) - \sum_{n=0}^{N-1} x_n \psi_n (t) \right|^2 dt = 0
\end{equation*}
when $x(t)$ is a deterministic signal, and
\begin{equation*}
\lim_{N \rightarrow \infty} E \left\{ \left| x(t) - \sum_{n=0}^{N-1} x_n \psi_n (t) \right|^2 \right\} = 0, \;t \in \mathcal{A}\text,
\end{equation*}
when $x(t)$ is a random process. Both of the statements are standard \cite{DavenportText}, and, for all expansions in orthogonal functions, the convergence in the mean-square sense is tacitly assumed in the remainder of this paper unless otherwise stated.} and
\begin{equation}
x_n = \frac{1}{\lambda_n} \int_{\mathcal A} x (t) \psi^*_n (t) dt \text.\label{EQ:KLCoeff}\\
\end{equation}
In addition, the coefficients of the expansion satisfies

\begin{equation}
E \{ x_n \} = 0 \text,
\end{equation}
and

\begin{equation}
E \left\{ x_{n_1} x^*_{n_2} \right\} = \delta_{n_1, n_2} \text.
\end{equation}

\label{THM:KarhunenLoeve}
\end{theorem}

Therefore, the Karhunen-Lo\'{e}ve (KL) expansion (\ref{EQ:KLOrthoExp}) provides an expansion of the random process in orthogonal basis functions with uncorrelated coefficients. For Gaussian random processes, Theorem \ref{THM:KarhunenLoeve} has stronger properties as stated in the following corollary.

\begin{corollary}
When $x(t) \in \mathcal{L}_{\mathcal{A}}^2$ is a zero-mean proper complex Gaussian random process with a strictly positive definite ACF $R(t,s)$, the convergence of the KL expansion (\ref{EQ:KLOrthoExp}) is almost sure\footnote{For every $\epsilon>0$,
\begin{equation*}
\Pr \left\{ \lim_{N \rightarrow \infty} \left| x(t) - \sum_{n=0}^{N-1} x_n \psi_n (t) \right| < \epsilon \right\} = 1 \text.
\end{equation*}} and the coefficients $x_n$ in (\ref{EQ:KLCoeff}) are independent and identically distributed (i.i.d.) circular-symmetric Gaussian random variables with zero-mean and unit variance. \label{COR:KarhunenLoeve}
\end{corollary}

Now the following theorem introduces the PSWFs, which, in conjunction with Theorem \ref{THM:KarhunenLoeve}, comprise the foundation for the spectral decomposition of the scattering response.

\begin{theorem} [Slepian and Pollak \cite{Slepian61}]
Consider a positive definite kernel on a compact interval ${\mathcal A} = [a,b]$,
\begin{equation}
R (t, s) = W \sinc \left( W (t-s) \right), \;(t,s) \in {\mathcal A}^2 \text,
\end{equation}
where $W > 0$. Let $\{\psi_n (t)\}$ and $\{\lambda_n\}$ be the sets of eigenfunctions and eigenvalues of $\mathcal{T}_{\mathcal{A}} (R)$, respectively. 
Then, the following properties hold.
\begin{enumerate}
\item $\psi_n (t)$ are real and bandlimited: $\psi_n (t) \in {\mathcal B}_{W}$.
\item $\psi_n (t)$ are orthonormal on the real line and complete in ${\mathcal B}_{W}$:
\begin{equation}
\int^{\infty}_{-\infty} \psi_{n_1} (t) \psi_{n_2} (t) dt = \delta_{n_1,n_2}, \;n_1, n_2 = 0,1,2,\ldots \text.
\end{equation}
\item $\psi_n (t)$ are orthogonal on ${\mathcal A}$ and complete in ${\mathcal L}^2_{\mathcal A}$:
\begin{equation}
\int_{\mathcal A} \psi_{n_1} (t) \psi_{n_2} (t) dt = \lambda_{n_1} \delta_{n_1,n_2}, \;n_1, n_2 = 0,1,2,\ldots \text.
\end{equation}
\item $\lambda_n$ are real, positive, and distinct: $1 > \lambda_0 > \lambda_1 > \lambda_2 > \cdots$. 
\item For $t \in \mathbb{R}$,
\begin{equation}
\int_{\mathcal A} R (t,s) \psi_n (s) ds = \lambda_n \psi_n (t), \;n=0,1,2,\ldots \text.
\end{equation}
\end{enumerate}
\label{THM:SlepianPollak}
\end{theorem}

The widely used terminology to refer to $\psi_n (t)$ in Theorem \ref{THM:SlepianPollak} is the prolate spheroidal wave functions. In the remainder of this paper, we denote $\left\{ \psi_n (t) \right\}$ by $\mathcal{P}_{\mathcal{A},W}$, where the parameters are explicitly indicated in the subscript. 

\begin{remark}
The compact interval ${\mathcal A}$ in Theorems \ref{THM:KarhunenLoeve} and \ref{THM:SlepianPollak} can be replaced by a finite union of disjoint compact intervals. The extension of Theorem \ref{THM:KarhunenLoeve} for the multiple intervals can be found in \cite{Ferreira09}. Also, the extension of Theorem \ref{THM:SlepianPollak} was initially pursued in \cite{Landau80} and the asymptotic distribution of the eigenvalues was derived. The composition of PSWFs for the multiple interval case can be found in \cite{Izu09, Khare06}.
\end{remark}

\subsection{Spectral decomposition of the scattering response} \label{SEC:ChDecomp}

For the transmit ACF $R_t (\alpha_1,\alpha_2)$, the sets of eigenfunctions and eigenvalues of $\mathcal{T}_{\Omega_t} (R_t)$ are denoted by $\left\{ \psi_{t,n} (\alpha) \right\}$ and $\left\{ \lambda_{t,n} \right\}$, respectively. 
Then, from (\ref{EQ:TxCorr}) and by Theorem \ref{THM:SlepianPollak}, $\psi_{t,n} (\alpha)$ are PSWFs. In multi-bounce diffuse scattering, $h (\beta, \alpha)$ has an ACF proportional to $R_t (\alpha_1,\alpha_2)$ on $\Omega_t$ for a fixed $\beta \in \Omega_r$. By Theorem \ref{THM:KarhunenLoeve}, it can be expanded as
\begin{equation}
h(\beta,\alpha) = \sum_{n=0}^{\infty} h_n (\beta) \psi_{t,n} (\alpha), \;\alpha \in \Omega_t \text, \label{EQ:TxKLExpan}
\end{equation}
where
\begin{equation}
h_n (\beta) = \frac{1}{\lambda_{t,n}} \int_{\Omega_t} h(\beta,\alpha) \psi_{t,n} (\alpha) d \alpha \text.
\end{equation}
Note that $h_n (\beta)$ is a Gaussian random process with
\begin{equation}
E \left\{ h_n (\beta) \right\} = 0 \text,
\end{equation}
and

\begin{align}
E \left\{ h_{n_1} (\beta_1) h^*_{n_2} (\beta_2) \right\} &= \frac{1}{\lambda_{t,n_1} \lambda_{t,n_2}} \int_{\Omega_t} \int_{\Omega_t} E\left\{ h(\beta_1,\alpha_1) h^* (\beta_2,\alpha_2) \right\} \psi_{t,n_1} (\alpha_1) \psi_{t,n_2} (\alpha_2) d \alpha_1 d \alpha_2 \nonumber\\
&\overset{\rm(a)}{=} \frac{1}{\lambda_{t,n_1} \lambda_{t,n_2}} \int_{\Omega_t} \int_{\Omega_t} R_r (\beta_1,\beta_2) R_t (\alpha_1,\alpha_2) \psi_{t,n_1} (\alpha_1) \psi_{t,n_2} (\alpha_2) d \alpha_1 d \alpha_2 \nonumber\\
&\overset{\rm(b)}{=} \frac{1}{\lambda_{t,n_1}} \int_{\Omega_t} R_r (\beta_1,\beta_2) \psi_{t,n_1} (\alpha_1) \psi_{t,n_2} (\alpha_1) d \alpha_1 \nonumber\\
&\overset{\rm(c)}{=} R_r (\beta_1,\beta_2) \cdot \delta_{n_1,n_2}\text, \label{EQ:hbetaACF}
\end{align}
where (a) is from the definition of the autocorrelation function (\ref{EQ:ColoredDiff}) and (b) and (c) are by the properties of PSWFs in Theorem \ref{THM:SlepianPollak}.

Now, associated with the receive ACF $R_r (\beta_1,\beta_2)$, we define $\left\{ \psi_{r,n} (\beta) \right\}$ and $\left\{ \lambda_{r,n} \right\}$ as the sets of eigenfunctions and eigenvalues of $\mathcal{T}_{\Omega_r} (R_r)$, respectively. Then, $\psi_{r,n} (\beta)$ are PSWFs from (\ref{EQ:RxCorr}) and by Theorem \ref{THM:SlepianPollak}. Since $h_n (\beta)$ has an ACF of $R_r (\beta_1, \beta_2)$ as shown in (\ref{EQ:hbetaACF}), it can be expanded by Theorem \ref{THM:KarhunenLoeve} as
\begin{equation}
h_n (\beta) = \sum_{m=0}^{\infty} h_{m,n} \psi_{r,m} (\beta), \;\beta \in \Omega_r \text, \label{EQ:RxKLExpan}
\end{equation}
where
\begin{equation}
h_{m,n} = \frac{1}{\lambda_{r,m}} \int_{\Omega_r} h_n (\beta) \psi_{r,m} (\beta) d \beta \text.
\end{equation}
Also, by Corollary \ref{COR:KarhunenLoeve}, $h_{m,n}$ are zero-mean i.i.d. circular symmetric Gaussian random variables, i.e.,
\begin{equation}
E \left\{ h_{m,n} \right\} = 0
\end{equation}
and
\begin{equation}
E \left\{ h_{m_1,n_1} h^*_{m_2,n_2} \right\} = \delta_{m_1,m_2} \delta_{n_1,n_2} \text.
\end{equation}
Therefore, (\ref{EQ:TxKLExpan}) and (\ref{EQ:RxKLExpan}) yield an expansion
\begin{equation}
h(\beta,\alpha) = \sum_{m=0}^{\infty} \sum_{n=0}^{\infty} h_{m,n} \psi_{r,m} (\beta) \psi_{t,n} (\alpha), \; (\beta, \alpha) \in \Omega_r \times \Omega_t \text. \label{EQ:ChDecomp}
\end{equation}
Note that the expansion (\ref{EQ:ChDecomp}) is intrinsic to the channel and does not depend on the transmit and receive arrays. The basis functions $\psi_{t,n} (\alpha)$ and $\psi_{r,n} (\beta)$ form the transmit and receive subspaces, respectively, along which the signal can be transmitted. 
However, as it will be seen in the following section, it depends on the arrays how those subspaces are exploited.




\section{Capacity bound for the colored scattering channel} \label{SEC:CapBound}

In this section, an upper bound for the ergodic capacity of the colored scattering channel is presented as a function of the array length, the channel correlation width, the total angular spread of scattering clusters, and the signal-to-noise ratio (SNR). Before that, we first consider a useful lemma regarding the relationship between two different sets of PSWFs. 

\begin{lemma} \label{LEM:PSWFSubset}
Consider a compact interval (or finite union of disjoint compact intervals) $\mathcal{A}$ and bandwidths $W_1$ and $W_2$, where $W_1 \geq W_2 > 0$. Let $\left\{ \psi_n (t) \right\}$ and $\left\{ \phi_n (t) \right\}$ be $\mathcal{P}_{\mathcal{A},W_1}$ and $\mathcal{P}_{\mathcal{A},W_2}$, respectively. The corresponding sets of eigenvalues for $\left\{ \psi_n (t) \right\}$ and $\left\{ \phi_n (t) \right\}$ are also denoted by $\left\{ \lambda_n \right\}$ and $\left\{ \gamma_n \right\}$, respectively. Then there exist real coefficients $c_{m,n}$ ($m,n=0,1,2,\ldots$) such that
\begin{equation}
\phi_m (t) = \sum_{n=0}^{\infty} c_{m,n} \psi_n (t) \label{EQ:Prop0_phi}
\end{equation}
and
\begin{equation}
\zeta_n (t) = \lambda_n \sum_{m=0}^{\infty} c_{m,n} \phi_m (t) \text,
\end{equation}
where
\begin{equation}
\zeta_{n}(t) \triangleq \int_{\mathcal{A}} W_2 \sinc \left( W_2 (t-s) \right) \psi_{n}(s) ds \text. \label{EQ:Prop0_zeta}
\end{equation}
In addition, the coefficients $c_{m,n}$ satisfies
\begin{equation}
\delta_{m_1,m_2} = \sum_{n=0}^{\infty} c_{m_1,n} c_{m_2,n} \text,
\end{equation}
and
\begin{equation}
\gamma_{m_1} \delta_{m_1,m_2} = \sum_{n=0}^{\infty} \lambda_n c_{m_1,n} c_{m_2,n} \text.
\end{equation}
\end{lemma}

\begin{IEEEproof}
Note that $\mathcal{B}_{W_2} \subseteq \mathcal{B}_{W_1}$ since $W_2 \leq W_1$. Therefore, by property 2 of Theorem \ref{THM:SlepianPollak}, we can decompose $\phi_m (t)$ in terms of the basis functions $\psi_n (t)$ as (\ref{EQ:Prop0_phi}), where
\begin{equation}
c_{m,n} = \frac{1}{\lambda_n} \int_{\mathcal{A}} \phi_m (t) \psi_n(t) dt \text. \label{EQ:Prop0_coeff}
\end{equation}
Since both $\phi_m (t)$ and $\psi_n (t)$ are real functions, the coefficients $c_{m,n}$ are also real. Then, from the orthogonality of $\phi_m (t)$, we have

\begin{align}
\int_{-\infty}^{\infty} \phi_{m_1} (t) \phi_{m_2} (t) dt &= \delta_{m_1,m_2} \nonumber\\
&= \sum_{n=0}^\infty c_{m_1,n} c_{m_2,n} \text,
\end{align}
where the second equality follows by using (\ref{EQ:Prop0_phi}) and the orthonormality of $\psi_n(t)$ on the real line. That is, the sequences $\left[ c_{m,0}, c_{m,1}, c_{m,2}, \ldots \right]$ are orthonormal for different $m$. In addition, we have

\begin{align}
\int_{\mathcal{A}} \phi_{m_1} (t) \phi_{m_2} (t) dt &= \gamma_{m_1} \delta_{m_1,m_2} \nonumber\\
&= \sum_{n=0}^\infty \lambda_n c_{m_1,n} c_{m_2,n} \text,
\end{align}
where the second equality follows by using (\ref{EQ:Prop0_phi}) and the orthogonality of $\psi_n(t)$ on $\mathcal{A}$. Now, from (\ref{EQ:Prop0_zeta}), note that $\zeta_n (t)$ are the $\mathcal{A}$-timelimited and $W_2$-bandlimited versions of $\psi_n (t)$. Then, since $\zeta_n (t)$ are $W_2$-bandlimited, they can be expressed as
\begin{equation}
\zeta_n (t) = \sum_{m=0}^{\infty} \bar{c}_{n,m} \phi_m (t) \text, \label{EQ:Prop1_zeta}
\end{equation}
where

\begin{align}
\bar{c}_{n,m} &= \frac{1}{\gamma_m} \int_{\mathcal{A}} \zeta_n (t) \phi_m(t) dt \nonumber\\
&\overset{(a)}{=} \frac{1}{\gamma_m} \int_{\mathcal{A}} \int_{\mathcal{A}} W_2 \sinc \left( W_2 (t-s) \right) \psi_n (s) ds \phi_m (t) dt \nonumber\\
&\overset{(b)}{=} \frac{1}{\gamma_m} \int_{\mathcal{A}} \int_{\mathcal{A}} W_2 \sinc \left( W_2 (s-t) \right) \phi_m (t) dt \psi_n (s) ds \nonumber\\
&\overset{(c)}{=} \int_{\mathcal{A}} \phi_m (s) \psi_n (s) ds\nonumber\\
&\overset{(d)}{=} \lambda_n c_{m,n} \text,
\end{align}
where (a) is from (\ref{EQ:Prop0_zeta}), (b) is by changing the order of integrations and by the even symmetry of the sinc function, (c) is by property 5 of Theorem \ref{THM:SlepianPollak}, and (d) is from (\ref{EQ:Prop0_coeff}).
%
%
%
\end{IEEEproof}

In the following subsections, discrete subchannel decompositions of the transmit and receive signals are discussed. Along with the spectral decomposition of the scattering response (\ref{EQ:ChDecomp}), an equivalent canonical representation of the relationship in (\ref{EQ:RxSigAngle}) is obtained in the angular domain and the upper bound for the ergodic capacity of the canonical model is investigated.

\subsection{Subchannel decomposition} \label{SEC:SubchDecomp}

We suppose that the transmit ACF $R_t(\alpha_1, \alpha_2)$ and $\Omega_t$ are known at the transmitter, and the receiver has full channel state information. In particular, for the sake of simplicity, we focus on the symmetric case in which we assume $L_t = L_r = L$, $\Gamma_t = \Gamma_r = \Gamma$, $M_t = M_r$, and $|\Omega_t| = |\Omega_r| = |\Omega|$, where $|\Omega| = \int_{\Omega} d\alpha$. Since the subchannel decomposition depends on the array length $L$ and the correlation width $\Gamma$, we first tackle the case with $L \leq 1/\Gamma$, and then the case with $L > 1/\Gamma$ afterward.

\subsubsection{Case 1: $L \leq 1/\Gamma$}

Denote the eigenfunction and eigenvalue sets of $\mathcal{T}_{\Omega_t} \left( Q \right)$, where $Q (\alpha_1,\alpha_2) = L \sinc \left( L (\alpha_1-\alpha_2) \right)$, by $\left\{ \phi_{t,n} (\alpha) \right\}$ and $\left\{ \gamma_{t,n} \right\}$, respectively. By Theorem \ref{THM:SlepianPollak}, the eigenfunctions form a PSWF set $\mathcal{P}_{\Omega_t,L}$, which is a basis for $\mathcal{B}_{L}$. Therefore, without loss of generality, any transmit signals from the array of length $L$ can be expressed in the angular domain as
\begin{equation}
X (\alpha) = \sum_{n=0}^{\infty} X_n \phi_{t,n} (\alpha) \text, \label{EQ:TxDecomp0}
\end{equation}
where
\begin{equation}
X_n = \frac{1}{\gamma_{t,n}} \int_{\Omega_t} X(\alpha) \phi_{t,n} (\alpha) d \alpha \text.
\end{equation}
By Parseval's theorem, the transmit power constraint (\ref{EQ:TxPowConst}) can be rewritten as

\begin{align}
E \left\{ \int_{-\infty}^{\infty} \left| X (\alpha) \right|^2 d \alpha \right\} &= \sum_{n=0}^{\infty} E \left\{ \left| X_n \right|^2 \right\} \nonumber\\
&\leq P \text. \label{EQ:PowerConst2}
\end{align}
Analogously, we denote the eigenfunction and eigenvalue sets of $\mathcal{T}_{\Omega_r} (Q)$ by $\left\{ \phi_{r,m} (\beta) \right\}$ and $\left\{ \gamma_{r,m} \right\}$, respectively. By Theorem \ref{THM:SlepianPollak}, $\left\{ \phi_{r,m} (\beta) \right\}$ is $\mathcal{P}_{\Omega_r, L}$ and forms a basis for $\mathcal{B}_{L}$. 
Therefore, any received signals of the array of length $L$ can be represented in the angular domain by
\begin{equation}
Y (\beta) = \sum_{m=0}^{\infty} Y_m \phi_{r,m} (\beta) \text, \label{EQ:RxDecomp0}
\end{equation}
where
\begin{equation}
Y_m = \frac{1}{\gamma_{r,m}} \int_{\Omega_r} Y (\beta) \phi_{r,m} (\beta) d \beta \text. \label{EQ:RxSigCoeff}
\end{equation}
In particular, the noise $Z (\beta)$ embedded in $Y (\beta)$ is decomposed as
\begin{equation}
Z (\beta) = \sum_{m=0}^{\infty} Z_m \phi_{r,m} (\beta) \text, \label{EQ:RxNoise0}
\end{equation}
where
\begin{equation}
Z_m = \frac{1}{\gamma_{r,m}} \int_{\Omega_r} Z (\beta) \phi_{r,m} (\beta) d \beta \text. \label{EQ:RxNoise}
\end{equation}
Then, it can easily be shown that $Z_m$ are i.i.d. circular symmetric Gaussian random variables with zero-mean and variance of $\sigma^2$:

\begin{align}
E \left\{ Z_{m_1} Z_{m_2}^* \right\} &= \frac{1}{\gamma_{r,m_1} \gamma_{r,m_2}} \int_{\Omega_r} \int_{\Omega_r} E \left\{ Z (\beta_1) Z (\beta_2) \right\} \phi_{r,m_1} (\beta_1) \phi_{r,m_2} (\beta_2) d \beta_1 d \beta_2 \nonumber\\
&\overset{(a)}{=} \frac{\sigma^2}{\gamma_{r,m_1} \gamma_{r,m_2}} \int_{\Omega_r} \int_{\Omega_r} Q (\beta_1,\beta_2) \phi_{r,m_1} (\beta_1) \phi_{r,m_2} (\beta_2) d \beta_1 d \beta_2\nonumber\\
&\overset{(b)}{=} \frac{\sigma^2}{\gamma_{r,m_1}} \int_{\Omega_r} \phi_{r,m_1} (\beta_1) \phi_{r,m_2} (\beta_1) d \beta_1\nonumber\\
&\overset{(c)}{=} \sigma^2 \delta_{m_1,m_2} \text,
\end{align}
where (a) is by the definition of $Z (\beta)$ in (\ref{EQ:NoiseAngle}), and (b) and (c) are by the properties of PSWFs in Theorem \ref{THM:SlepianPollak}.

Now, we substitute (\ref{EQ:ChDecomp}) into (\ref{EQ:RxSigAngle}) and obtain

\begin{align}
Y (\beta) &= \int_{\Omega_r} \int_{\Omega_t} Q (\beta,\beta') \sum_{m=0}^{\infty} \sum_{n=0}^{\infty} h_{m,n} \psi_{r,m} (\beta') \psi_{t,n} (\alpha') \int_{-\infty}^{\infty} Q (\alpha',\alpha) X (\alpha) d \alpha d \alpha' d \beta' + Z(\beta) \nonumber\\
&= \sum_{m=0}^{\infty} \sum_{n=0}^{\infty} h_{m,n} \zeta_{r,m} (\beta) \int_{-\infty}^{\infty} \zeta_{t,n} (\alpha) X(\alpha) d \alpha + Z(\beta) \text, \label{EQ:RxSigAngle_1}
\end{align}
where

\begin{align}
\zeta_{t,n} (\alpha) \triangleq \int_{\Omega_t} Q (\alpha',\alpha) \psi_{t,n} (\alpha') d \alpha' \label{EQ:Zeta_t_0} \text,\\
\zeta_{r,m} (\beta) \triangleq \int_{\Omega_r} Q (\beta,\beta') \psi_{r,m} (\beta') d \beta' \label{EQ:Zeta_r_0}\text.
\end{align}
From the assumption that $L \leq 1/\Gamma$ and by Lemma \ref{LEM:PSWFSubset}, (\ref{EQ:Zeta_t_0}) and (\ref{EQ:Zeta_r_0}) can be expressed as

\begin{align}
\zeta_{t,n} (\alpha) = \lambda_{t,n} \sum_{k=0}^{\infty} c_{k,n} \phi_{t,k} (\alpha) \label{EQ:Zeta_t_1} \text,\\
\zeta_{r,m} (\beta) = \lambda_{r,m} \sum_{l=0}^{\infty} d_{l,m} \phi_{r,l} (\beta) \label{EQ:Zeta_r_1} \text,
\end{align}
for some real coefficients $c_{k,n}$ and $d_{l,m}$. Plugging (\ref{EQ:TxDecomp0}), (\ref{EQ:RxNoise0}), (\ref{EQ:Zeta_t_1}), and (\ref{EQ:Zeta_r_1}) into (\ref{EQ:RxSigAngle_1}) and using (\ref{EQ:RxSigCoeff}) and (\ref{EQ:RxNoise}), we have a discrete canonical representation of the received signal
\begin{equation}
Y_l = \sum_{m=0}^{\infty} \sum_{n=0}^{\infty}  d_{l,m} \lambda_{r,m} h_{m,n} \lambda_{t,n} \sum_{k=0}^{\infty} c_{k,n} X_k + Z_l,\; l=0,1,2,\ldots \text. \label{EQ:RxSigCase1}
\end{equation}

\subsubsection{Case 2: $L > 1/\Gamma$}

As discussed in the previous case, any transmit signals from the array of length $L$ can be expressed in terms of basis functions $\phi_{t,n} (\alpha)$ as (\ref{EQ:TxDecomp0}). From the spectral decomposition of the scattering response (\ref{EQ:ChDecomp}), the subspace in the angular domain, through which signals can be conveyed from the transmitter to the receiver, is spanned by $\left\{ \psi_{t,n} (\alpha) \right\}$ on $\Omega_t$. By extending the support of the basis functions from $\Omega_t$ to the real line and by Theorem \ref{THM:SlepianPollak}, we see that any signals that can be conveyed through the channel can be interpolated to signals in $\mathcal{B}_{1/\Gamma}$. In other words, $X(\alpha) \in \mathcal{B}_{1/\Gamma}$ is a necessary condition for signals not to be nulled by the channel. 
On the other hand, the subspace spanned by $\left\{ \phi_{t,n} (\alpha) \right\}$ on the real line is $\mathcal{B}_{L}$, which includes $\mathcal{B}_{1/\Gamma}$ when $L > 1/\Gamma$. 
Therefore, it is sufficient for the transmitter to use $\left\{ \psi_{t,n} (\alpha) \right\}$ as the basis for the transmit signal instead of $\left\{ \phi_{t,n} (\alpha) \right\}$. The transmit signal is now given by
\begin{equation}
X (\alpha) = \sum_{n=0}^{\infty} X_n \psi_{t,n} (\alpha) \text, \label{EQ:TxDecomp1}
\end{equation}
where
\begin{equation}
X_n = \frac{1}{\lambda_{t,n}} \int_{\Omega_t} X(\alpha) \psi_{t,n} (\alpha) d \alpha \text.
\end{equation}
Note that the same power constraint as (\ref{EQ:PowerConst2}) is applied in this case. Using (\ref{EQ:ChDecomp}) and (\ref{EQ:TxDecomp1}) in (\ref{EQ:RxSigAngle}), we have
\begin{equation}
Y (\beta) = \sum_{m=0}^{\infty} \sum_{n=0}^{\infty} h_{m,n} \int_{\Omega_r} Q (\beta,\beta') \psi_{r,m} (\beta') d \beta' \lambda_{t,n} X_n + Z(\beta) \text, \label{EQ:RxSigAngle_2}
\end{equation}
where the fact that $X(\alpha)$ in (\ref{EQ:TxDecomp1}) is $1/\Gamma$-bandlimited, i.e., $\int_{-\infty}^{\infty} Q (\alpha',\alpha) X (\alpha) d \alpha = X (\alpha')$, is used. By Lemma \ref{LEM:PSWFSubset}, when $L > 1/\Gamma$, we can write
\begin{equation}
\psi_{r,m} (\beta) = \sum_{l=0}^{\infty} \bar{d}_{m,l} \phi_{r,l} (\beta)
\end{equation}
for some real coefficients $\bar{d}_{m,l}$. Therefore,
\begin{equation}
\int_{\Omega_r} Q(\beta,\beta') \psi_{r,m} (\beta') d \beta' = \sum_{l=0}^{\infty} \gamma_{r,l} \bar{d}_{m,l} \phi_{r,l} (\beta) \text. \label{EQ:psi}
\end{equation}
As in the previous case, any received signals from the array of length $L$ can be expressed as (\ref{EQ:RxDecomp0}). Using (\ref{EQ:RxSigAngle_2}) and (\ref{EQ:psi}) in (\ref{EQ:RxDecomp0}) and (\ref{EQ:RxSigCoeff}), we obtain
\begin{equation}
Y_l = \sum_{m=0}^{\infty} \sum_{n=0}^{\infty} \gamma_{r,l} \bar{d}_{m,l} h_{m,n} \lambda_{t,n} X_n + Z_l,\; l=0,1,2,\ldots \text, \label{EQ:RxSigCase2}
\end{equation}
where $Z_l$ are i.i.d. circular symmetric Gaussian random variables with zero-mean and variance of $\sigma^2$ as defined in (\ref{EQ:RxNoise}).

\subsection{capacity upper bound}

The capacity upper bounds for the canonical representations (\ref{EQ:RxSigCase1}) and (\ref{EQ:RxSigCase2}) are investigated. For the characterization of the capacity, the distributions of eigenvalues $\lambda_{t,n}$ or $\gamma_{t,n}$, and $\lambda_{r,m}$ or $\gamma_{r,m}$ are essential, since they measure the gains of transmit and receive subchannels. In the earlier work by Landau and Widom \cite{Landau80}, the asymptotic distribution of those eigenvalues was found as in the following theorem.

\begin{theorem} [Landau and Widom \cite{Landau80}]
Let $\mathcal{A}$ be a union of $M$ ($< \infty$) disjoint compact intervals. Consider a kernel on $\mathcal{A}$
\begin{equation}
R (t, s) = W \sinc \left( W (t-s) \right), \;(t,s) \in {\mathcal A}^2 \text,
\end{equation}
where $W > 0$. Then the number of eigenvalues of $\mathcal{T}_{\mathcal{A}} (R)$ that exceed $x$ ($0<x<1$) is given by
\begin{equation}
\bar{G}_{\mathcal{A},W} (x) = \left| \mathcal{A} \right| W + \frac{M}{\pi^2} \ln \frac{1-x}{x} \ln \left( 2 \pi \left| \mathcal{A} \right| W \right) + o \left( \ln \left( \left| \mathcal{A} \right| W \right) \right),\; \left| \mathcal{A} \right| W \rightarrow \infty \text,
\end{equation}
where $| \mathcal{A} | = \int_{\mathcal{A}} dt$. \label{THM:EvDist}
\end{theorem}

\begin{figure} [t]
\begin{center}
\epsfig{file=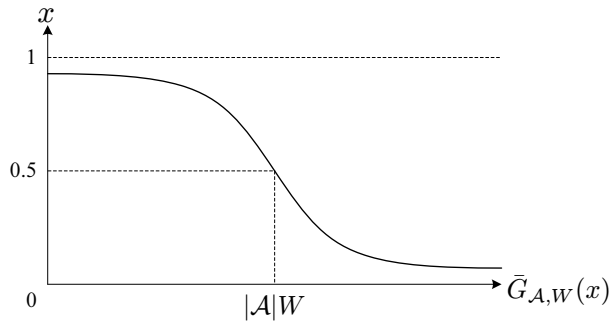, width=8cm}
\caption{Distribution of eigenvalues of $\mathcal{T}_{\mathcal{A}} (R)$ in Theorem \ref{THM:EvDist}.} \label{FIG:EvDistrib}
\end{center}
\end{figure}

The distribution of the eigenvalues is graphically shown in Fig. \ref{FIG:EvDistrib}.

Now we first consider the upper bound for the capacity when $L \leq 1/\Gamma$. For notational simplicity, let us consider an infinite-dimensional vector space representation of (\ref{EQ:RxSigCase1}) as
\begin{equation}
{\bf y} = {\bf D} {\bf \Lambda}_r {\bf H} {\bf \Lambda}_t {\bf C}^T {\bf x} + {\bf z} \label{EQ:RxSignalDiscrete} \text,
\end{equation}
where ${\bf y} = [Y_1, Y_2, \ldots]^T$, ${\bf x} = [X_1, X_2, \ldots]^T$, and ${\bf z} = [Z_1, Z_2, \ldots]^T$. Also, ${\bf D}$, ${\bf H}$, and ${\bf C}$ are matrices whose $(m,n)$-th elements are given by $d_{m,n}$, $h_{m,n}$, and $c_{m,n}$, respectively, and ${\bf \Lambda}_r$ and ${\bf \Lambda}_t$ are diagonal matrices whose $n$-th diagonal elements are $\lambda_{r,n}$ and $\lambda_{t,n}$, respectively. It is well-known that the capacity of the system (\ref{EQ:RxSignalDiscrete}) is achieved by a zero-mean Gaussian input ${\bf x}$ and given by
\begin{equation}
C = \max_{\genfrac{}{}{0pt}{}{{\bf Q}:}{{{\rm Tr}({\bf Q}) \leq P}}} E \left\{ \log \det \left[ {\bf I} + \frac{1}{\sigma^2} {\bf D} {\bf \Lambda}_r {\bf H} {\bf \Lambda}_t {\bf C}^T {\bf Q} {\bf C} {\bf \Lambda}_t {\bf H}^H {\bf \Lambda}_r {\bf D}^T \right] \right\} \label{EQ:Cap1} \text,
\end{equation}
where ${\bf Q} \triangleq E \left\{ {\bf x} {\bf x}^H \right\}$. Note that, due to the infinite dimensionality of the model, the determinant in (\ref{EQ:Cap1}) should be carefully defined in the limiting sense as in \cite{Chuah02, Wallace02, Poon06}. That is, we first assume that only finite numbers of dimensions of ${\bf x}$ and ${\bf y}$ in (\ref{EQ:RxSignalDiscrete}) are used, which gives the capacity representation (\ref{EQ:Cap1}), and then the numbers of dimensions are taken to infinity. Letting $\bar{\bf x} \triangleq {\bf C}^T {\bf x}$, we can see in (\ref{EQ:RxSignalDiscrete}) that ${\bf x}-\bar{\bf x}-{\bf y}$ forms a Markov chain. Also, due to the orthonormality of the coefficients $c_{m,n}$ (Lemma \ref{LEM:PSWFSubset}), $\bar{\bf x} = [\bar{X}_0, \bar{X}_1, \ldots]^T$ satisfies the same power constraint:
\begin{equation}
\sum_{n=0}^{\infty} E \left\{ \left| \bar{X}_n \right|^2 \right\} = \sum_{k=0}^{\infty} E \left\{ \left| X_k \right|^2 \right\} \leq P \text. \label{EQ:PowerConst3}
\end{equation}
Therefore, we have the following upper bound for the capacity:
\begin{equation}
C \leq \max_{\genfrac{}{}{0pt}{}{\bar{\bf Q}:}{{{\rm Tr}(\bar{\bf Q}) \leq P}}} E \left\{ \log \det \left[ {\bf I} + \frac{1}{\sigma^2} {\bf D} {\bf \Lambda}_r {\bf H} {\bf \Lambda}_t \bar{\bf Q} {\bf \Lambda}_t {\bf H}^H {\bf \Lambda}_r {\bf D}^T \right] \right\} \label{EQ:Cap2} \text,
\end{equation}
where $\bar{\bf Q} \triangleq E \left\{ \bar{\bf x} \bar{\bf x}^H \right\}$. In \cite{Telatar95}, it was shown that the optimal input covariance matrix to achieve the ergodic capacity is a scaled identity matrix when the channel matrix has i.i.d. entries. Though this is not the case in (\ref{EQ:Cap2}), it was shown in \cite{Veeravalli05, Jafar01, Jorswieck04} that a diagonal input covariance matrix achieves the ergodic capacity if any two column vectors of the channel matrix are independent and the entries have a symmetric distribution around zero. As a result, without loss of generality, we can let $\bar{\bf Q}$ in (\ref{EQ:Cap2}) be a diagonal matrix with $E \{ | \bar{X}_n |^2 \}$ on its diagonal. Based on the fact that $\log \det (\cdot)$ is concave, Jensen's inequality is applied to (\ref{EQ:Cap2}) giving another upper bound

\begin{align}
C &\leq \max_{\genfrac{}{}{0pt}{}{\bar{\bf Q}:}{{{\rm Tr}(\bar{\bf Q}) \leq P}}} \log \det \left[ {\bf I} + \frac{1}{\sigma^2} {\bf D} {\bf \Lambda}_r E \left\{ {\bf H} {\bf \Lambda}_t \bar{\bf Q} {\bf \Lambda}_t {\bf H}^H \right\} {\bf \Lambda}_r {\bf D}^T \right]  \nonumber\\
&\overset{(a)}{=} \max_{\sum_{n=0}^{\infty} E \{ | \bar{X}_n |^2 \} \leq P} \log \det \left[ {\bf I} + \frac{\sum_{n=0}^{\infty} \lambda_{t,n}^2 E \left\{ \left| \bar{X}_n \right|^2 \right\} }{\sigma^2} {\bf D} {\bf \Lambda}_r^2 {\bf D}^T \right] \nonumber\\
&\overset{(b)}{\leq} \log \det \left[ {\bf I} + \frac{P}{\sigma^2} {\bf D} {\bf \Lambda}_r^2 {\bf D}^T \right] \label{EQ:Cap3} \text,
\end{align}
where (a) follows from the fact that $\bar{\bf Q}$ and ${\bf \Lambda}_t$ are diagonal and $h_{m,n}$ are i.i.d. random variables with zero-mean and unit variance, and (b) from the fact that $0< \lambda_{t,n} < 1$. By further applying Hadamard's inequality to (\ref{EQ:Cap3}), we obtain

\begin{align}
C &\leq \sum_{l=0}^{\infty} \log \left( 1+ \frac{P}{\sigma^2} \sum_{m=0}^{\infty} \lambda_{r,m}^2 d_{l,m}^2 \right) \nonumber\\
&\overset{(a)}{\leq} \sum_{l=0}^{\infty} \log \left( 1+ \frac{P}{\sigma^2} \sum_{m=0}^{\infty} \lambda_{r,m} d_{l,m}^2 \right) \nonumber\\
&\overset{(b)}{=} \sum_{l=0}^{\infty} \log \left( 1+ \frac{P}{\sigma^2} \gamma_{r,l} \right) \text, \label{EQ:Cap4}
\end{align}
where (a) is from the fact that $0< \lambda_{r,m} < 1$ and (b) is by Lemma \ref{LEM:PSWFSubset}.

\begin{figure} [t]
\begin{center}
\epsfig{file=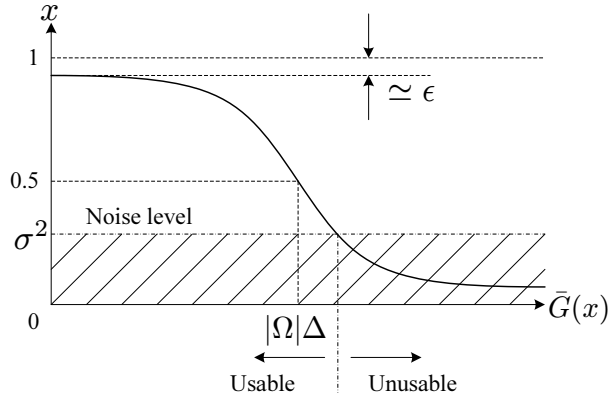, width=8cm}
\caption{Distribution of eigenvalues $\bar{G} (x)$.} \label{FIG:EvDistrib2}
\end{center}
\end{figure}

Now, we turn to the upper bound for $L > 1/\Gamma$ case. Analogously to the previous case, the infinite-dimensional vector space representation of (\ref{EQ:RxSigCase2}) is given by
\begin{equation}
{\bf y} = \bar{\bf \Lambda}_r \bar{\bf D} {\bf H} {\bf \Lambda}_t {\bf x} + {\bf z} \text, \label{EQ:RxSignalDiscrete1}
\end{equation}
where $\bar{\bf D}$ is a matrix whose $(m,n)$-th element is $\bar{d}_{m,n}$, $\bar{\bf \Lambda}_r$ is a diagonal matrix whose $n$-th diagonal element is $\gamma_{r,n}$, and the other variables are defined as in (\ref{EQ:RxSignalDiscrete}). The capacity of the system (\ref{EQ:RxSignalDiscrete1}) is then characterized as
\begin{equation}
C = \max_{\genfrac{}{}{0pt}{}{{\bf Q}:}{{{\rm Tr}({\bf Q}) \leq P}}} E \left\{ \log \det \left[ {\bf I} + \frac{1}{\sigma^2} \bar{\bf \Lambda}_r \bar{\bf D} {\bf H} {\bf \Lambda}_t {\bf Q} {\bf \Lambda}_t {\bf H}^H \bar{\bf D}^T \bar{\bf \Lambda}_r \right] \right\} \label{EQ:Cap8} \text.
\end{equation}
We apply the same bounding procedure as in (\ref{EQ:Cap3}) to (\ref{EQ:Cap8}), which yields

\begin{align}
C &\leq \log \det \left[ {\bf I} + \frac{P}{\sigma^2} \bar{\bf \Lambda}_r \bar{\bf D}^T \bar{\bf D} \bar{\bf \Lambda}_r \right] \nonumber\\
&\overset{(a)}{=} \log \det \left[ {\bf I} + \frac{P}{\sigma^2} \bar{\bf D} \bar{\bf \Lambda}_r^2 \bar{\bf D}^T \right] \label{EQ:Cap9} \text,
\end{align}
where (a) is by Sylvester's determinant theorem. Applying Hadamard's inequality to (\ref{EQ:Cap9}), we have

\begin{align}
C &\leq \sum_{m=0}^{\infty} \log \left( 1+ \frac{P}{\sigma^2} \sum_{l=0}^{\infty} \gamma_{r,l}^2 \bar{d}_{m,l}^2 \right) \nonumber\\
&\overset{(a)}{\leq} \sum_{m=0}^{\infty} \log \left( 1+ \frac{P}{\sigma^2} \sum_{l=0}^{\infty} \gamma_{r,l} \bar{d}_{m,l}^2 \right) \nonumber\\
&\overset{(b)}{=} \sum_{m=0}^{\infty} \log \left( 1+ \frac{P}{\sigma^2} \lambda_{r,m} \right) \text, \label{EQ:Cap10}
\end{align}
where (a) is from the fact that $0 < \gamma_{r,l} < 1$ and (b) is by Lemma \ref{LEM:PSWFSubset}.
Note that the upper bounds (\ref{EQ:Cap4}) and (\ref{EQ:Cap10}) are determined by the distribution of eigenvalues $\gamma_{r,l}$ and $\lambda_{r,m}$. Whether $L \leq 1/\Gamma$ or $L > 1/\Gamma$, the number of eigenvalues $\gamma_{r,l}$ or $\lambda_{r,m}$ exceeding $x$ ($0<x<1$) can collectively be quantified by Theorem \ref{THM:EvDist} as
\begin{equation}
\bar{G} (x) = \left| \Omega \right| \Delta + \frac{M}{\pi^2} \ln \frac{1-x}{x} \ln \left( 2 \pi \left| \Omega \right| \Delta \right) + o \left( \ln \left( \left| \Omega \right| \Delta \right) \right),\;\left| \Omega \right| \Delta \rightarrow \infty \text, \label{EQ:EvDist1}
\end{equation}
where $\Delta \triangleq \min \left\{ L, 1/\Gamma \right\}$. Define
\begin{equation}
G (x) = -\frac{M}{\pi^2} \ln \frac{1-x}{x} \ln \left( 2\pi \left| \Omega \right| \Delta \right) \label{EQ:EvDist2}
\end{equation}
such that $\bar{G} (x) = -G (x) + \left| \Omega \right| \Delta + o \left( \ln \left( \left| \Omega \right| \Delta \right) \right)$. Also, a parameter $\epsilon$ is determined to satisfy
\begin{equation}
\left| \Omega \right| \Delta + \frac{M}{\pi^2} \ln \frac{\epsilon}{1-\epsilon} \ln \left( 2 \pi \left| \Omega \right| \Delta \right) = 0 \text.
\end{equation}
Then, as shown in Fig. \ref{FIG:EvDistrib2}, $1-\epsilon$ approximately represents the largest eigenvalue. Using the collective distribution of eigenvalues (\ref{EQ:EvDist1}), the upper bounds in (\ref{EQ:Cap4}) and (\ref{EQ:Cap10}) can be written in a single expression as
\begin{equation}
C \leq \int_0^{1-\epsilon} \log \left( 1+ \frac{x P}{\sigma^2} \right) d G(x) + o \left( \ln \left( \left| \Omega \right| \Delta \right) \right),\;\left| \Omega \right| \Delta \rightarrow \infty \text. \label{EQ:Cap6}
\end{equation}
The same form of integration on the right-hand-side of (\ref{EQ:Cap6}) has already been dealt with in \cite[Lemma 3.4]{Poon06}, and (\ref{EQ:Cap6}) can be compactly reformulated as
\begin{equation}
C \leq \left[ \left| \Omega \right| \Delta + M  \ln \left( 2 \pi \left| \Omega \right| \Delta \right) f \left( \frac{P}{\sigma^2} \right) \right] \log \left( 1+ \frac{P}{\sigma^2} \right) + o \left( \ln \left( \left| \Omega \right| \Delta \right) \right),\;\left| \Omega \right| \Delta \rightarrow \infty \text, \label{EQ:Cap7}
\end{equation}
where
\begin{equation}
f \left( \frac{P}{\sigma^2} \right) = \frac{1}{2 \pi^2} \ln \frac{P}{\sigma^2} + o \left( \ln \frac{P}{\sigma^2} \right),\;\frac{P}{\sigma^2} \rightarrow \infty \text.
\end{equation}
One interesting point to observe in (\ref{EQ:Cap7}) is that the pre-log factor, i.e., the number of degrees of freedom, is dependent on the SNR. In particular, the dependence becomes prominent when $| \Omega | \Delta$ is small. When arrays with discrete and finite number of antennas are considered, such dependence on SNR is not observed in general. In fact, the SNR-dependence of the number of degrees of freedom has already been pointed out multiple times in the literature considering continuous arrays \cite{Bucci89, Migliore06, Poon06}. It can intuitively be explained as follows. As shown in (\ref{EQ:Cap4}) and (\ref{EQ:Cap10}), the eigenvalues of the array or the channel represent gains of parallel channels. Therefore, assuming unit signal power, $P=1$, the number of eigenvalues exceeding the noise level will determine the number of usable\footnote{This means that the eigenvalues above the noise level have dominant contributions to the overall capacity. It does not imply that the capacity vanishes when $\sigma^2 >1$ because no channel is usable.} channels. By Theorem \ref{THM:EvDist}, about $| \Omega | \Delta$ eigenvalues are close to $1$ and then plunge near zero in a transition region of width around $M \ln ( 2 \pi |\Omega| \Delta )/\pi^2$. With reference to Fig. \ref{FIG:EvDistrib2}, if the transition of eigenvalues from $\simeq 1$ to $\simeq 0$ is not abrupt enough compared to $|\Omega| \Delta$, i.e., for a small $|\Omega| \Delta$ value, the number of usable channels will noticeably depend on the noise level.

The result of this section is summarized in the following theorem.

\begin{theorem}
For the channel under colored scattering with correlation width $\Gamma$ and total angular spread $| \Omega |$, the ergodic capacity achieved by linear transmit and receive arrays of length $L$ at a given SNR $P/\sigma^2$ is bounded by
\begin{equation}
C \leq |\Omega| \Delta \log \left(1+ \frac{P}{\sigma^2} \right) + o \left( |\Omega| \Delta \right)
\end{equation}
as $| \Omega | \Delta \rightarrow \infty$, where $\Delta \triangleq \min \left\{ L, 1/\Gamma \right\}$. \label{THM:Capacity}
\end{theorem}

Interestingly, and as somewhat expected, the result in Theorem \ref{THM:Capacity} coincides with the result in \cite[Theorem 3.5]{Poon06} when $L \leq 1/\Gamma$. That is, at a given SNR, the capacity may almost linearly scale in proportion to the array length $L$. However, in the colored scattering environment, the capacity gets saturated to an intrinsic limit posed by the channel.

\section{Numerical results} \label{SEC:NumericalResult}

To validate the theoretic result provided in Section \ref{SEC:CapBound}, some simulations are performed. Since the continuous scattering and continuous array models in Section \ref{SEC:ScatteringModel} are not convenient for numerical simulation, we consider the discrete scattering and discrete array models as in \cite{Sayeed02}. Assuming linear arrays with $2L+1$ discrete, half-wavelength-spaced antennas (indexed from $-L$ to $L$), the discretized version of the channel response (\ref{EQ:ChResp}) between the $m$-th receive and $n$-th transmit antennas is given by
\begin{equation}
\tilde{c}_{m,n} = \eta \sum_{k=-K}^K \sum_{l=-K}^K \tilde{a}_r^* (k,m) \tilde{h} (k,l) \tilde{a}_t (l,n),\; m,n=-L,\ldots,L \text, \label{EQ:DiscreteChCoeff}
\end{equation}
for some nonnegative integer $K$, where $\eta$ is a scaling factor and

\begin{align}
\tilde{a}_t (l,n) &= e^{-j 2\pi \frac{l}{K} \frac{n}{2}},\; l=-K,\ldots,K, \; n=-L,\ldots,L \text,\\
\tilde{a}_r (k,m) &= e^{-j 2\pi \frac{k}{K} \frac{m}{2}},\; k=-K,\ldots,K, \; m=-L,\ldots,L \text.
\end{align}
In addition, to capture the correlation between scattered fields, the discrete scattering response $\tilde{h} (k,l)$ is randomly generated to have a circular symmetric Gaussian distribution with zero-mean and an ACF
\begin{equation}
E \left\{ \tilde{h}(k_1, l_1) \tilde{h}^* (k_2, l_2) \right\} = \frac{1}{\Gamma^2} \sinc \left( \frac{k_1 - k_2}{K \Gamma} \right) \sinc \left( \frac{l_1 - l_2}{K \Gamma} \right) \text,
\end{equation}
for
\begin{equation}
\left(\frac{k_1}{K},\frac{l_1}{K}\right), \left(\frac{k_2}{K},\frac{l_2}{K}\right) \in \Omega^2 \text.
\end{equation}
More specifically, we choose $K=2^{11}$, $L=0,1,\ldots,49$, and $\Omega = [-1.0, -0.7] \cup [-0.15, 0.15] \cup [0.7, 1.0]$, i.e., $M=3$ and $|\Omega|=0.9$. Also, the scaling factor $\eta$ in (\ref{EQ:DiscreteChCoeff}) is determined such that $\sum_{m=-L'}^{L'} \sum_{n=-L'}^{L'} E \left\{ \left| \tilde{c}_{m,n} \right|^2 \right\}/(2L'+1)^2 = 1$ for $L'=49$. Then, the resultant discrete antenna signal model is given by
\begin{equation}
\tilde{\bf y} = \tilde{\bf H} \tilde{\bf x} + \tilde{\bf z} \text,
\end{equation}
where $\tilde{\bf x}$ and $\tilde{\bf y}$ are $(2L+1) \times 1$ transmitted and received vectors, $\tilde{\bf z}$ is a $(2L+1) \times 1$ noise vector comprised of i.i.d. circular symmetric Gaussian elements with zero-mean and variance of $\sigma^2$, and $\tilde{\bf H}$ is a $(2L+1) \times (2L+1)$ channel matrix whose $(m,n)$-th element ($m,n=-L,\ldots,L$) is $\tilde{c}_{m,n}$. According to the assumption in Section \ref{SEC:SubchDecomp}, the transmitter only knows the distribution of the channel; correlation between entries of $\tilde{\bf H}$. 
Based on this information and the transmit power constraint $P$, the optimal input covariance matrix $\tilde{\bf Q} \triangleq E \left\{ \tilde{\bf x} \tilde{\bf x}^H \right\}$ is determined to achieve the ergodic capacity. However, determining the optimal input covariance matrix and computing the ergodic capacity of the correlated MIMO channel is cumbersome, because there is no closed form expression and, thus, numerical optimization is required \cite{Jafar01, Jorswieck04}.
Therefore, as alternatives, we evaluate two performance metrics in \cite{Chuah02}, which serve as the lower and upper bounds for the ergodic capacity, respectively: the first one is the average mutual information with equal power allocation, given by
\begin{equation}
\tilde{I}_{2L+1} = E \left\{ \log \det \left[ {\bf I} + \frac{P}{\sigma^2 (2L+1)} \tilde{\bf H} \tilde{\bf H}^H \right] \right\} \label{EQ:ExpMutualInfo} \text,
\end{equation}
and the second one is the average capacity with full channel state information at both transmitter and receiver, given by
\begin{equation}
\tilde{C}_{2L+1} = E \left\{ \max_{\genfrac{}{}{0pt}{}{\tilde{\bf Q}:}{{{\rm Tr}(\tilde{\bf Q}) \leq P}}} \log \det \left[ {\bf I} + \frac{1}{\sigma^2} \tilde{\bf H} \tilde{\bf Q} \tilde{\bf H}^H \right] \right\} \label{EQ:ExpCapacity} \text.
\end{equation}

\begin{figure} [t]
\begin{center}
\subfigure[$P/\sigma^2 = 0$dB]{\epsfig{file=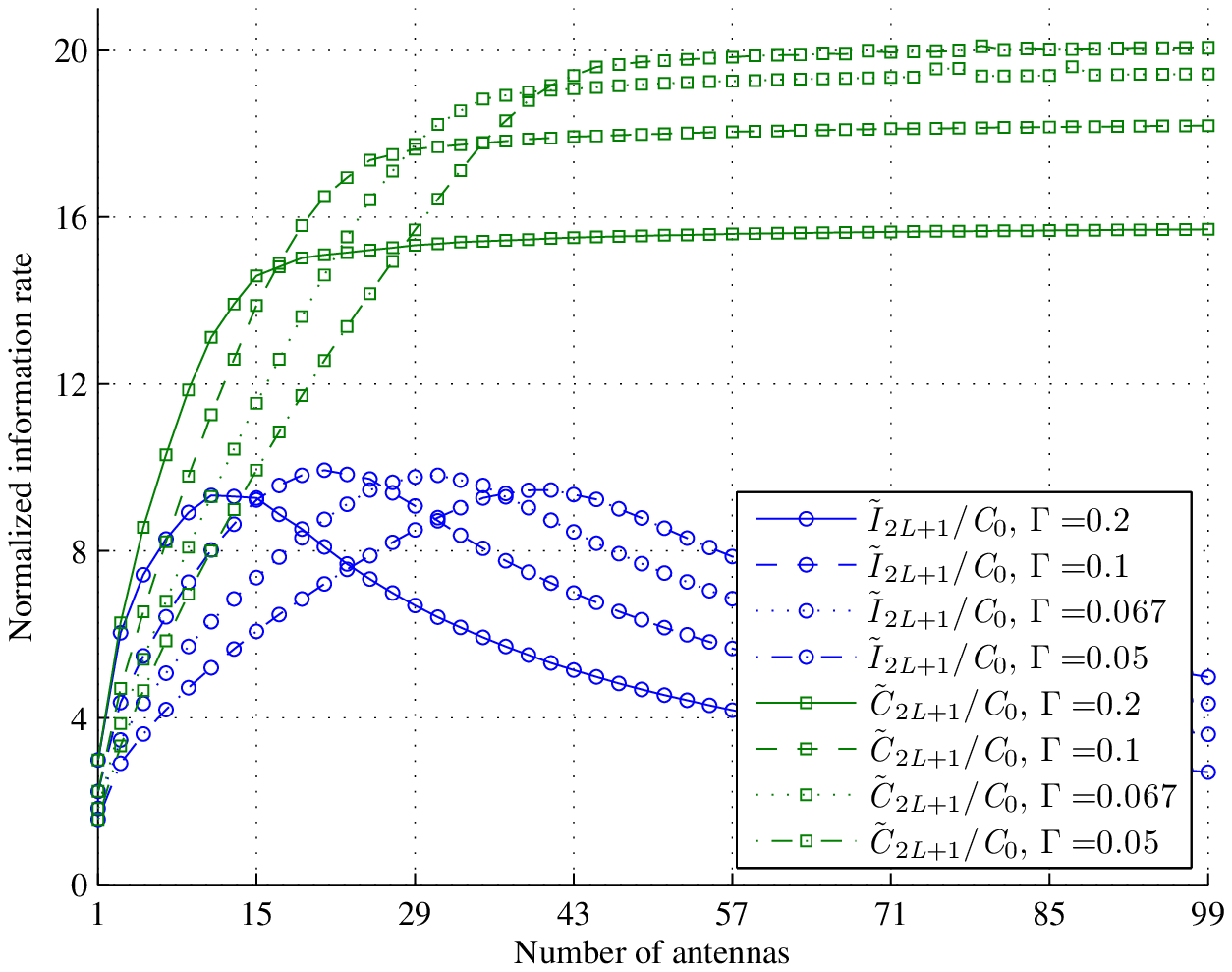,
width=0.49\linewidth}\label{FIG:DOF00dB}} \subfigure[$P/\sigma^2 = 15$dB]{\epsfig{file=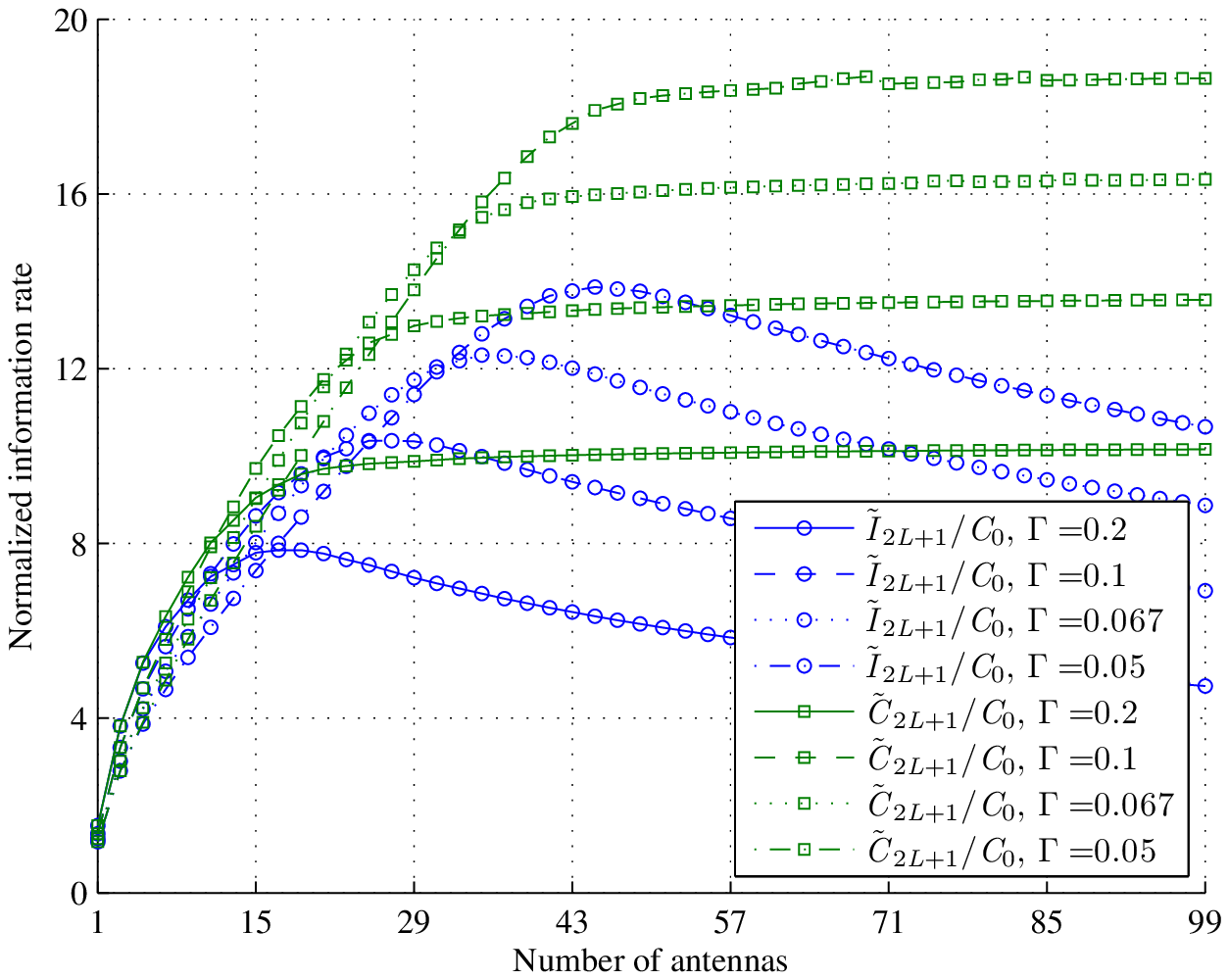,
width=0.49\linewidth}\label{FIG:DOF15dB}} \subfigure[$P/\sigma^2 = 30$dB]{\epsfig{file=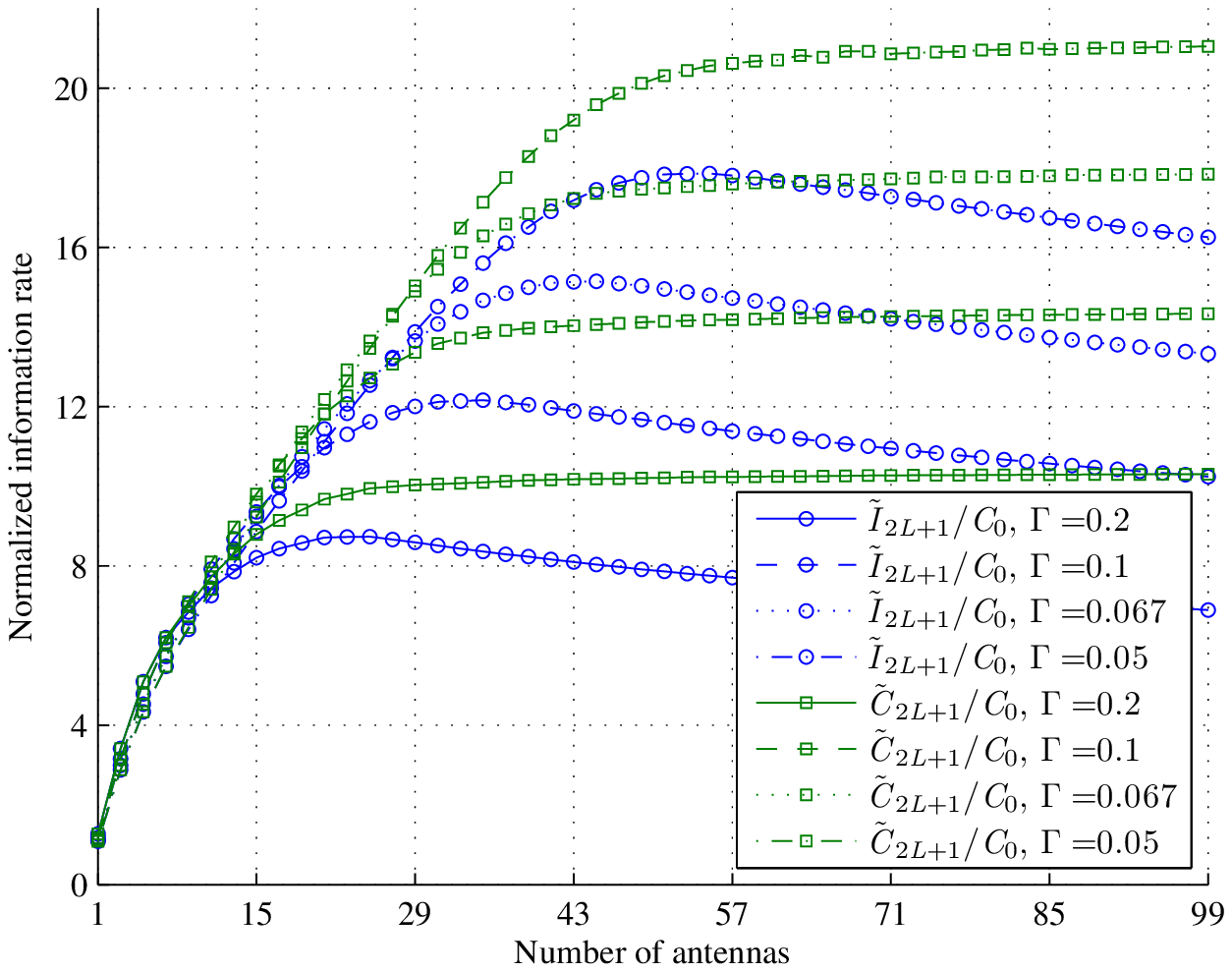,
width=0.49\linewidth}\label{FIG:DOF30dB}} \subfigure[$P/\sigma^2 = 45$dB]{\epsfig{file=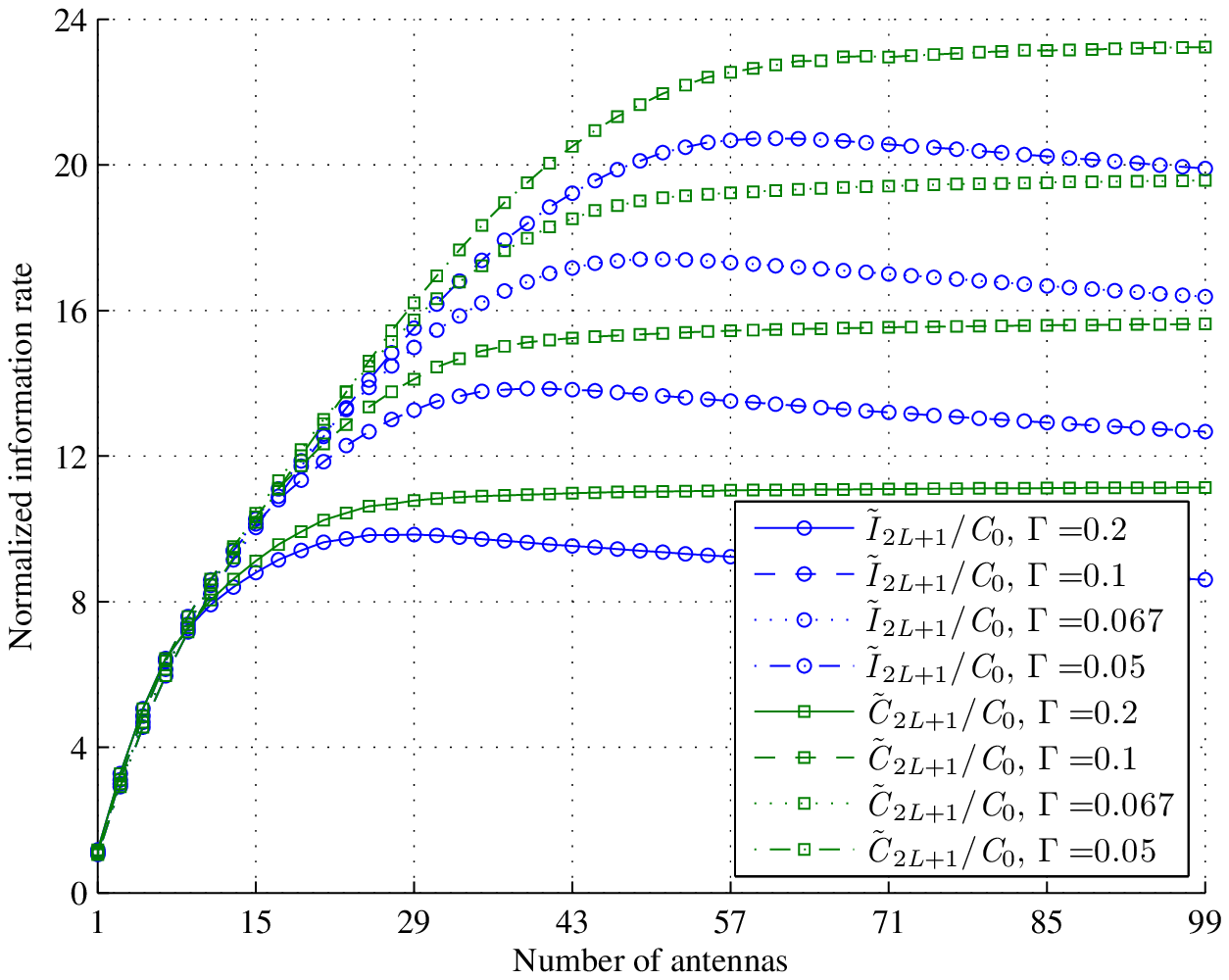,
width=0.49\linewidth}\label{FIG:DOF45dB}} \caption{Bounds for the ergodic capacity of the colored scattering channel; linear arrays with half-wavelength-spaced antennas are assumed.}
\label{FIG:DOF}
\end{center}
\end{figure}

The values of $\tilde{I}_{2L+1}$ and $\tilde{C}_{2L+1}$ are computed through Monte Carlo simulation with $10^4$ trials, and the curves of $\tilde{I}_{2L+1}$ and $\tilde{C}_{2L+1}$, normalized by $C_0 = \log ( 1+ P/\sigma^2 )$, are plotted in Fig. \ref{FIG:DOF} for different values of $\Gamma$ and different SNRs. Note that, since the inter-antenna spacing is fixed to half a wavelength, the sizes of array apertures grow with the number of antennas. In the figure, we can observe that the growth rates of $\tilde{I}_{2L+1}$ and $\tilde{C}_{2L+1}$ with respect to the number of antennas (equivalently, array aperture size) are severely affected by the correlation width $\Gamma$. It is seen that they do not linearly increase with the number of antennas and, in particular, the upper bound $\tilde{C}_{2L+1}$ eventually gets saturated to a certain value; the larger the correlation width, to the smaller value the upper bound gets saturated. On the other hand, the lower bound $\tilde{I}_{2L+1}$ increases with the number of antennas up to a certain point and then gradually decreases. This is because the available degrees of freedom are not efficiently used, while waterfilling is used for the upper bound to make the best use of the degrees of freedom. However, the gap between the upper and lower bounds narrows as the SNR increases. In summary, the results in Fig. \ref{FIG:DOF} verify the statement in Section \ref{SEC:CapBound}: in the colored scattering environment, the ergodic capacity may not linearly grow with the array size but gets saturated to a certain limit.




%
%
%
%
%
%

\section{Discussion} \label{SEC:Discussion}

\subsection{Single-bounce diffuse scattering}

In the analyses in Sections \ref{SEC:ChDecomp} and \ref{SEC:CapBound}, only the multi-bounce diffuse channel has been considered. As noted in Section \ref{SEC:ScatteringModel}, single- and multi-bounce diffuse channels share the same ACF (\ref{EQ:ColoredDiff}) but have different supports. Both in the single- and multi-bounce cases, the ACF can be expressed as a product of the transmit and receive ACFs as in (\ref{EQ:ACFFact}). However, in the single-bounce case, the transmit and receive ACFs are not completely separable in the sense that the supports of those two functions are intertwined. If we consider a single scattering cluster at once and ignore the other clusters, the transmit and receive ACFs are separable; the spectral decomposition of the scattering response in the pair of transmit and receive subintervals corresponding to each cluster is given by
\begin{equation}
h(\beta,\alpha) = \sum_{m=0}^{\infty} \sum_{n=0}^{\infty} h^{(i)}_{m,n} \psi^{(i)}_{r,m} (\beta) \psi^{(i)}_{t,n} (\alpha), \; (\beta, \alpha) \in \Omega_{r,i} \times \Omega_{t,i}, i=1,2,\dots,M \text, \label{EQ:ChDecompSingle}
\end{equation}
where $h^{(i)}_{m,n}$ are i.i.d. circular symmetric Gaussian random variables with zero-mean and unit variance, and $\psi^{(i)}_{t,n}(\alpha)$ and $\psi^{(i)}_{r,m}(\beta)$ are PSWFs $\mathcal{P}_{\Omega_{t,i},1/\Gamma}$ and $\mathcal{P}_{\Omega_{r,i},1/\Gamma}$, respectively. Assuming $L \leq 1/\Gamma$, the transmit and receive signal can be written similarly to the multi-bounce counterpart as
\begin{equation}
X(\alpha) = \sum_{i=1}^M \sum_{n=0}^{\infty} X^{(i)}_n \phi^{(i)}_{t,n} (\alpha)
\end{equation}
and
\begin{equation}
Y(\beta) = \sum_{i=1}^M \sum_{m=0}^{\infty} Y^{(i)}_m \phi^{(i)}_{r,m} (\beta) \text,
\end{equation}
where $\phi^{(i)}_{t,n}(\alpha)$ and $\phi^{(i)}_{r,m}(\beta)$ are PSWFs $\mathcal{P}_{\Omega_{t,i},L}$ and $\mathcal{P}_{\Omega_{r,i},L}$, respectively. In \cite{Landau80, Izu09}, the asymptotic orthogonality of the PSWFs with disjoint time or frequency supports was addressed. That is, for $i_1 \neq i_2$, $\psi^{(i_1)}_{t,n} (\alpha)$ and $\psi^{(i_2)}_{t,n} (\alpha)$ are asymptotically orthogonal either on $\Omega_{t,i_1}$ or on $\Omega_{t,i_2}$ for a small correlation width $\Gamma$ (large $1/\Gamma$). The same asymptotic orthogonality can be stated for $\psi^{(i)}_{r,m}(\beta)$, $\phi^{(i)}_{t,n}(\alpha)$, and $\phi^{(i)}_{r,m}(\beta)$ (large $L$). Due to this asymptotic orthogonality, the single-bounce diffuse channel is asymptotically equivalent to $M$ independent single-cluster multi-bounce diffuse channels (the same argument was made in \cite{Poon06}). Therefore, after some straightforward manipulation on (\ref{EQ:Cap7}), we have the capacity upper bound for the single-bounce diffuse channel as
\begin{equation}
C \leq \left[ \left| \Omega \right| \Delta + \sum_{i=1}^M  \ln \left( 2 \pi \left| \Omega_i \right| \Delta \right) f \left( \frac{P}{M \sigma^2} \right) \right] \log \left( 1+ \frac{P}{M \sigma^2} \right) + o \left( \ln \left( \left| \Omega \right| \Delta \right) \right),\;\left| \Omega_i \right| \Delta \rightarrow \infty \text, \label{EQ:CapSingle}
\end{equation}
where it is assumed that $|\Omega_{t,i}| = |\Omega_{r,i}| = |\Omega_i|$ and equal power is allocated for each cluster ($\sum_{n=0}^{\infty} E \{ | X_n^{(i)} |^2 \} \leq P/M$, $i=1,\ldots,M$). Comparing (\ref{EQ:Cap7}) and (\ref{EQ:CapSingle}), it is seen that the single- and multi-bounce diffuse channels have almost the same number of degrees of freedom for large $|\Omega| \Delta$.

%
%
%

\subsection{Diversity gain}

As for the diversity gain of the slow-fading colored scattering channel, the same argument for the white scattering channel in \cite{Poon06} can be revisited. Though the multi- and single-bounce diffuse channels have almost the same degrees of freedom, their diversity gains and the tradeoffs between the diversity and multiplexing gains are different. In particular, one thing to note for colored scattering is that, as in the case of degrees of freedom, the maximum diversity gain is limited by the correlation characteristic of the channel regardless of the lengths of arrays. By the same argument in \cite{Poon06}, it is clear that the limits are given by $|\Omega_t||\Omega_r|/(\Gamma_t \Gamma_r)$ for the multi-bounce case and $\sum_{i=1}^M |\Omega_{t,i}||\Omega_{r,i}|/(\Gamma_t \Gamma_r)$ for the single-bounce case, respectively.

\subsection{Conclusion}

The correlation in the scattered fields has not been captured in conventional scattering channel models. In this paper, we introduced a correlated scattering model, referred to as the colored scattering model, and analyzed the impact of correlation on the channel capacity. In previous studies, it has been well known that the capacity and the diversity gain grow in proportion to the sizes of transmit and receive array apertures under the conventional white scattering model. In contrast, it was shown in this paper that the capacity and diversity gain can be saturated to certain values under the colored scattering model. The limits on the capacity and diversity gain are intrinsically determined by the correlation characteristic of the channel and, as the correlation decreases, i.e., as the model gets close to the white scattering model, the limits increase indefinitely. That is, the colored scattering model includes the conventional white scattering model as the limiting case. The result of this paper yields useful insight into the interaction between the channel and antenna arrays and is applicable in various situations concerning very large arrays. For example, as the larger and larger antenna arrays are being taken into consideration in modern and future communication systems to attain high spectral efficiency and reliability, the result of this paper can provide a guideline for the selection of the maximum aperture size with the given correlation characteristic of the scattering environment.

\end{document}